\documentclass[10pt,aps,prd,twocolumn,superscriptaddress,showpacs,showkeys,nofootinbib,floatfix,preprintnumbers]{revtex4}
\usepackage{appendix}
\usepackage{epsfig}
\usepackage{graphicx}
\usepackage{amsfonts}
\usepackage{amsmath}
\usepackage{psfrag}
\usepackage{amssymb,amsmath,amsthm}
\usepackage{verbatim}
\usepackage{soul}
\usepackage[normalem]{ulem}
\usepackage{color}
\usepackage{multirow}
\usepackage{titlesec}

\newcommand{\beq}{\begin{equation}}
\newcommand{\eeq}{\end{equation}}
\newcommand{\bea}{\begin{eqnarray}}
\newcommand{\eea}{\end{eqnarray}}

\begin{document}

\title{Kinetic equation  approach to graphene in strong external fields}

% Author Orchid ID: enter ID or remove command
\iffalse
\newcommand{\orcidauthorA}{0000-0003-0611-4765} % Add \orcidA{} behind the author's name
\newcommand{\orcidauthorB}{0000-0003-2332-0982} % Add \orcidB{} behind the author's name
\newcommand{\orcidauthorC}{0000-0002-8399-5183} % Add \orcidC{} behind the author's name
\newcommand{\orcidauthorD}{0000-0003-0082-052X} % Add \orcidD{} behind the author's name
\fi
% Authors, for the paper (add full first names)
\author{Stanislav A. Smolyansky} 
\affiliation{Saratov State University, RU - 410026 Saratov, Russia}
\affiliation{Tomsk State University, RU - 634050 Tomsk, Russia}
\author{Anatolii D. Panferov}
\affiliation{Saratov State University, RU - 410026 Saratov, Russia}
\author{David B. Blaschke}
\affiliation{Institute of Theoretical Physics, University of Wroclaw, PL-50204 Wroclaw, Poland}
\affiliation{Bogoliubov Laboratory of Theoretical Physics, Joint Institute for Nuclear Research, 141980 Dubna, Russia}
\affiliation{National Research Nuclear University, 115409 Moscow, Russia}
\author{Narine T. Gevorgyan}
\affiliation{A.I. Alikhanyan National Science Laboratory (YerPhI), %Alikhanyan Brothers street 2, 
		0036 Yerevan, Armenia}

\begin{abstract}
	The report presents the results of using the nonperturbative kinetic approach to describe the excitation of plasma oscillations in a graphene monolayer.
	As examples the constant electric field as well as an electric field of short high-frequency pulses are considered. The dependence of the induced conduction and polarization currents characteristics on the pulse intensity, pulse duration, and polarization is investigated.
	The characteristics of secondary electromagnetic radiation resulting from the alternating currents is investigated. The nonlinear response to the external electric field characterizes graphene as an active medium.
	Qualitative agreement is obtained with the existing experimental result of measurements of currents in constant electric fields and radiation from graphene in the case of excitation by means of the infrared and optical pulses.
\end{abstract}
% Keywords
% The fields PACS, MSC, and JEL may be left empty or commented out if not applicable
\maketitle

\section{Introduction}
Graphene is a unique real material, promising for microelectronics and described by a fairly simple quantum field model (massless $D = 2 + 1$ QED). These features of graphene and the possibility of the experimental verification of the theoretical predictions explain the great interest in theoretical studies in the electrodynamics of graphene. A specifics here is the lack of analyticity in the coupling constant. This makes the use of nonperturbative approaches relevant here.

One of these directions is associated with the adaptation of general methods of QED for strong fields based on exact solutions of the main QED equations for some simple models of the external electric field \cite{Ritus, Nikishov, Fradkin:1991zq}. In graphene, this approach allows a detailed study of the electrodynamics in the case of a constant \cite{Gavrilov:2012jk} and pulsed (Sauter) electric field \cite{Klimchitskaya:2013fpa}. For the case of a constant field, satisfactory agreement with experiment was obtained \cite{Vandecasteele:2010, Kane:2015, Panferov:2019}.

Another recently proposed line of research in graphene is based on the use of methods of nonperturbative kinetic theory, which in the simplest case is valid for spatially homogeneous nonstationary electromagnetic fields. Within the framework of standard QED, such an approach was proposed in several papers \cite{Grib:1988, Bialynicky-Birula:1991, Schmidt:1998, Tarakanov:2002} and is currently successfully applied in various problems of QED in strong external fields (for example,  \cite{Blaschke:2009, Blaschke:2016}) in describing the vacuum production and evolution of an electron – positron plasma. A promising field of application of this approach is also in QCD, where problems of the creation and evolution of a quark-gluon plasma in the initial stage of the collision of ultrarelativistic heavy ions can be considered (for a review of the early works in this direction, see \cite{Vinnik:2001}). This {universal} approach was recently adapted to the model of single-layer graphene in \cite{Panferov:2017, Smolyansky:2019, Smolyansky_school:2019}.

The work consists of two interdependent parts.
Sections~\ref{sect:ps}--\ref{sect:ma} contain an account of  nonperturbative kinetic theory and electrodynamics of graphene based on the quasiparticle representation.
On this foundation the analysis of currents and plasma oscillations is performed in Sections~\ref{sect:cf}--\ref{sect:pf} for the cases of a constant and pulsed external fields.
These results are compared with the existing experimental data.
Finally, some actual perspectives on forthcoming investigations are discussed in Sections~\ref{sect:c}.

\section{Statement of the Problem \label{sect:ps}}
\label{main}

The aim of this work is the construction and subsequent study of the extended kinetic theory of carriers and radiation in single-layer plane graphene excited by the action of an external semiclassical field $A_{ex}^\mu$. Internal currents in turn generate internal fields $A_{in}^\mu$. Before making more realistic approximations, the quasiclassical fields are assumed to be spatially homogeneous, arbitrarily depending on time and acting in the graphene plane, so that in the Hamiltonian gauge $A^0=0$ the structure of the acting field is as follows:

\begin{equation}
A_{ex,in}^{\mu} (t)=\left( 0,A_{ex,in}^1 (t),A_{ex,in}^2 (t),0\right) .
\label{A1}
\end{equation}

These limitations already mean that the radiation of graphene at the frequencies of plasma oscillations is an isolated problem related to the extension of $D2 \to D3$ electrodynamics. Quasiclassical fields can be strong and are nonperturbatively taken into account, while possible quantum fields are described in the framework of the relevant perturbation theory and can be spatially inhomogeneous and leave the limits of the graphene plane.
Thus, the effective fields in graphene will be equal $(k=1,2)$

\begin{equation}
A^{k}(t)=A^k_{ex}(t)+A^k_{in}(t).
\label{A3}
\end{equation}

In the nonperturbative part of the problem, the methods and terminology of the kinetic theory based on the quasiparticle representation are used. 
Supplementing this system with the Maxwell equation for the internal field, we obtain a closed self-consistent system of equations describing at the kinetic level the dynamics of carriers and fields in graphene.

Below we consider the simplest low-energy model of graphene in the presence of an external field~(\ref{A1}), which describes the excitations in the vicinity of one of the two Dirac points at the boundaries of the Brillouin zone \cite{Novoselov:2005, Geim:2007, Castro:2009}. A generalization of the formalism to a tight-binding model can be found in~\cite{Smolyansky:2019}.

We write the equation of motion and the Hamiltonian of the basic model in an effective semiclassical field  $A^k(t)$  (\ref{A3})
\begin{eqnarray}
i\hbar \dot{\Psi}(\vec{x},t)= {\mathrm v}_F\hat{\vec{P}}\vec{\sigma}\Psi(\vec{x},t) .
\label{H5}
\end{eqnarray}

\begin{eqnarray}
H(t)= {\mathrm v}_F \int{d^2x\Psi^{\dagger}(\vec{x},t)\hat{\vec{P}}\vec{\sigma}\Psi(\vec{x},t)} ,
\label{H6}
\end{eqnarray}
where $\hat{P}_k=-i \hbar \nabla_k-(e/c) A_k(t)$ is the quasimomentum $(k=1,2)$, $\sigma_k$ are the Pauli matrices,
\begin{eqnarray}
\sigma_1 = 
\begin{pmatrix}
0 & 1 \\
1 & 0
\end{pmatrix},~~
\sigma_2 = 
\begin{pmatrix}
0 & -i \\
i & 0
\end{pmatrix},
\label{PauliM}
\end{eqnarray}
and ${\mathrm v}_F=10^6$ m/s is the Fermi velocity. The charge of an electron is $-e,e>0$. The wave function in (5), (6) is a two-component spinor.

The corresponding current density is equal
\begin{eqnarray}
j^k(\vec{x},t)=e {\mathrm v}_F \Psi^{\dagger} (\vec{x},t) \sigma^k \Psi (\vec{x},t) .
\label{CurDens}
\end{eqnarray}

\section{Transition to the Quasiparticle Representation  \label{sect:tqr}}

Standard quantum field theory is based on the transition to the representation of second quantization in momentum space, which allows for the usual physical interpretation, where the key element is the concept of particles, real or virtual, which are considered as excitations of the physical vacuum. As a result, in the case of free fields, all necessary quantities become state-additive in momentum space with a certain population determined by the statistics of the fields.

The introduction of interaction with an external field violates the additivity of the observables in the Fock space and complicates the interpretation of the formalism in terms of particles and~antiparticles.

The transition to the {quasiparticle} representation \cite{Grib:1988} to a certain extent allows us to solve this problem and restore clarity in the description of processes in strong fields.

In graphene, the entire scheme for constructing the kinetic theory of vacuum particle production in the standard QED in the quasiparticle representation is preserved, but it is much simplified when constructing the dynamics when implementing the canonical Bogolyubov transformation. The reason for the simplification is the absence of a mass and the reduction in the number of spatial dimensions. This leads to the fact that the spin degrees of freedom are hidden and degenerate. Their existence is reflected only in the flavor number $N_s=2$. The inclusion of real spin degrees of freedom requires a $D2 \to D3$ generalization of the theory.

The transition to the quasiparticle representation in graphene was used, for example, \mbox{in~\cite{Klimchitskaya:2013fpa, Panferov:2017, Smolyansky:2019, Dora:2010}}. Below we follow the works \cite{Klimchitskaya:2013fpa,  Panferov:2017, Smolyansky:2019}.

We suppose that graphene is located in a region bounded by a square with side $L$ and go to the momentum representation,
\begin{eqnarray}
\Psi(\vec{x},t)=\frac{1}{L}\displaystyle\sum_{\vec{p}} \Psi(\vec{p},t) e^{i\vec{p}\vec{x}/\hbar} 
\label{MoRep}
\end{eqnarray}

In this representation, we write the basic Hamiltonian  (\ref{H6})
\begin{eqnarray}
H(t)={\mathrm v}_F\frac{1}{L^2}\displaystyle\sum_{\vec{p}} \Psi^{\dagger}(\vec{p},t) \vec{P}\vec{\sigma}\Psi^{\dagger}(\vec{p},t) 
\label{Ht}
\end{eqnarray}
where $\vec{P}=\vec{p}-\frac{e}{c}\vec{A}(t)$ is the quasi-momentum, and the equation of motion (\ref{H5}) is
\begin{eqnarray}
i \hbar \dot{\Psi}\left( \vec{p},t\right) = {\mathrm v}_{F}\hat{\vec{P}}%
\vec{\sigma}\Psi \left( \vec{p},t\right) ,  
\label{Dirac}
\end{eqnarray}

The diagonalization of the Hamiltonian (\ref{Ht}) can be done explicitly using the unitary transformation  $\Psi=\mathcal{U}\Phi$ with the matrix \cite{Dora:2010}
\begin{eqnarray}
\mathcal{U}(t)=\frac{1}{\sqrt{2}}\left(
\begin{array}{cc}
\exp (-i \varkappa /2) & \exp (-i \varkappa /2) \\
\exp (i \varkappa /2) & -\exp (i \varkappa /2)
\end{array}
\right) .
\label{unitar}
\end{eqnarray}

The parameter  $\tan \varkappa=P^2/P^1$   is fixed here by the equality
\begin{eqnarray}
\mathcal{U}^{\dag }(t) {\mathrm v}_{F}\vec{P}\vec{\sigma}\mathcal{U}(t)=\varepsilon (\vec{p},t)\sigma _{3} = H_{\vec{p}}(t), 
\label{energyunitar}
\end{eqnarray}
where $\varepsilon (\vec{p},t)$ is the quasienergy
\begin{eqnarray}
\varepsilon(\vec{p},t)={\mathrm v}_{F}\sqrt{P^{2}}= {\mathrm v}_{F}\sqrt{(P^{1})^{2}+(P^{2})^{2}}.
\label{energy}
\end{eqnarray}

According to the selected low-energy model, the dispersion law (\ref{energy}) is valid in the vicinity of the Dirac point $p^2=0$ at the boundaries of the Brillouin zone.

The equation of motion in the quasiparticle representation will have the form
\begin{eqnarray}
i \hbar \dot{\Phi} =  H_{\vec{p}}(t) \Phi + \frac {1}{2}\lambda \hbar \sigma_1 \Phi ,  
\label{Dirac_t}
\end{eqnarray}
where the function
\begin{eqnarray}  \label{lambda}
\lambda \left(\vec{p},t\right)=\dot{\varkappa}=\frac{e \mathrm v_{F}^{2}[E_{1}P_{2}-E_{2}P_{1}]}{\varepsilon^{2}(\vec{p},t)}.
\end{eqnarray}
describes the transition between states with positive and negative energies and can be found from the~equality
\begin{eqnarray}
2i\mathcal{U}^{\dagger}\dot{\mathcal{U}}=\lambda\sigma_1 .  
\label{UU}
\end{eqnarray}
In the formula (\ref{lambda}) $E_k (t)=-\frac{1}{c} \dot{A}_k(t)$ is the electric field strength.

If we now write the spinor $\Phi(\vec{p},t)$ taking into account expansion (\ref{MoRep}) in the form
\begin{eqnarray}
\Phi (\vec{p},t)=\left[
\begin{array}{c}
a(\vec{p},t) \\
b^{\dag }(-\vec{p},t)
\end{array}
\right] , 
\label{components}
\end{eqnarray}
then the Hamiltonian of the system in the new representation will be equal to:
\begin{eqnarray}
H(t)&=&\frac{1}{L^2}\sum_{\vec{p}}\varepsilon(\vec{p},t)\Phi^\dagger(\vec{p},t)\sigma_3\Phi(\vec{p},t)\nonumber\\
&=&\frac{1}{L^2}\sum_{\vec{p}}\varepsilon(\vec{p},t)\left[a^{\dagger} a(\vec{p},t)-b(-\vec{p},t)b^\dagger(-\vec{p},t) \right]. \nonumber\\
\label{Hsystem}
\end{eqnarray}

The spinor (\ref{components}) can be associated with the field operator
\begin{eqnarray}
\Phi (\vec{x},t)=\frac{1}{L}\sum_{\vec{p}}\left\lbrace a(\vec{p},t)\hat{u}+b^\dagger(-\vec{p},t)\hat{v} \right\rbrace e^{i\vec{p}\vec{x}/\hbar} , 
\label{FieldOp}
\end{eqnarray}
where
\begin{eqnarray}
\hat{u} =\left[
\begin{array}{c}
1 \\
0
\end{array}
\right] , ~~
\hat{v} =\left[
\begin{array}{c}
0 \\
1
\end{array}
\right] 
\label{ss}
\end{eqnarray}
are the unit spinors.

We now turn to the representation of occupation numbers and define the creation and annihilation operators of electrons and holes over the instantaneous vacuum state $\left| t\right\rangle $, demanding for these operators the standard anticommutation relations,
\begin{eqnarray}
\left\lbrace a(\vec{p},t),a^\dagger(\vec{p}',t) \right\rbrace _+=\left\lbrace b(\vec{p},t),b^\dagger(\vec{p}',t) \right\rbrace _+=\delta_{\vec{p}\vec{p}'}.
\label{ar}
\end{eqnarray}
The remaining elementary anticommutators are equal to zero.

The equations of motion (\ref{Dirac_t}) can be written using (\ref{components}) in terms of amplitudes
\begin{eqnarray}
i\hbar \dot{a}(\vec{p},t)&=&\varepsilon (\vec{p},t) a(\vec{p},t)-\frac{i}{2}\hbar\lambda(\vec{p},t)b^\dagger(-\vec{p},t),
\nonumber \\
i\hbar \dot{b}(-\vec{p},t)&=&\varepsilon (\vec{p},t) b(-\vec{p},t)+\frac{i}{2}\hbar\lambda(\vec{p},t)a^\dagger(-\vec{p},t)
\label{ampl}
\end{eqnarray}
or in the form of Heisenberg-type equations with respect to the operator $\vartheta$
\begin{eqnarray}
i\hbar\dot{\vartheta}=[\vartheta,H_f(t)],
\label{Heis}
\end{eqnarray}
where
\begin{eqnarray}
H_f(t)=H(t)+H_{pol}(t) 
\label{Hferm}
\end{eqnarray}
is the Hamiltonian of the fermion subsystem, including the Hamiltonian $H(t)$ (\ref{Hsystem}) and the Hamiltonian

\begin{eqnarray}
H_{pol}(t)=-i\frac{\hbar}{2L^2}\sum_{\vec{p}}\lambda(\vec{p},t)&\left[ a^\dagger(\vec{p},t)b^\dagger(-\vec{p},t)\right.\nonumber\\
&\left.-b(-\vec{p},t)a(\vec{p},t)\right] 
\label{Hpol}
\end{eqnarray}
describing the effects of vacuum polarization under the influence of the field.

\section{Basic Kinetic Equation  \label{sect:bke}}

Below, in the quasiparticle representation, under kinetic equation (KE) will be understood the closed integro-differential equations for the distribution functions of electrons and holes
\begin{eqnarray}  
\label{DF}
f^{e}(\vec{p},t)&=&\langle {\rm in} |a^{+}(\vec{p},t)a(\vec{p},t)|{\rm in}\rangle, \\
f^{h}(\vec{p},t)&=&\langle {\rm in} |b^{+}(-\vec{p},t)b(-\vec{p},t)|{\rm in}\rangle.  %\notag
\label{DF2}
\end{eqnarray}
Subsequently, these functions are considered equal,
\begin{eqnarray}
f^e(\vec{p},t)=f^h(\vec{p},t)=f(\vec{p},t) \, ,
\label{fe}
\end{eqnarray}
by virtue of the assumption of electroneutrality of the system at each moment of time. 
The averaging in (\ref{DF}), (\ref{DF2}) is performed over the in-vacuum state.

To get a closed system of KE, we differentiate $f(\vec{p},t)$ with respect to time and use the equations of motion (\ref{ampl})
\begin{eqnarray}  
\label{DFequat}
\dot{f}(\vec{p},t)= \frac{1}{2} \lambda(\vec{p},t)\left[  f^{(+)}(\vec{p},t)+f^{(-)}(\vec{p},t) \right] ,
\end{eqnarray}
where anomalous averages are introduced by
\begin{eqnarray}  
\label{DFanomal}
f^{(+)}(\vec{p},t)&=&\langle {\rm in}| a^{+}(\vec{p},t)b^{+}(-\vec{p},t) |{\rm in}\rangle, \\
f^{(-)}(\vec{p},t)&=&\langle {\rm in}| b(-\vec{p},t)a(\vec{p},t) |{\rm in}\rangle~.  %\notag
\label{DFanomal2}
\end{eqnarray}
Differentiating them in time, we obtain
{\small
	\begin{eqnarray}  
	\label{DFanomalequat}
	\dot{f}^{(+)}(\vec{p},t)=\frac{2 i}{\hbar}\varepsilon(\vec{p},t)f^{(+)}(\vec{p},t) - \frac {\lambda(\vec{p},t)}{2}[1-2f(\vec{p},t)] \\
	\dot{f}^{(-)}(\vec{p},t)=\frac{-2 i}{\hbar}\varepsilon(\vec{p},t)f^{(-)}(\vec{p},t) +\frac {\lambda(\vec{p},t)}{2}[1-2f(\vec{p},t)].
	\label{DFanomalequat2}
	\end{eqnarray}
}

As a result of the integration of these equations over time and substitution in (\ref{DFequat}), we obtain the desired KE of non-Markovian type
\begin{eqnarray}
\dot{f}(\vec{p},t)=\frac{1}{2}\lambda \left( \vec{p},t\right)
\int\limits_{t_{0}}^{t}dt^{\prime }\lambda (\vec{p},t^{\prime })\left[ 1-2f(\vec{p},t^{\prime })\right] \cos \theta (t,t^{\prime }),  \nonumber\\
\label{KE}
\end{eqnarray}%
where the dynamic phase is introduced by
\begin{eqnarray}
\theta (t,t^{\prime })=\frac{2}{\hbar }\int\limits_{t^{\prime}}^{t}dt^{\prime \prime }\varepsilon (\vec{p},t^{\prime \prime }).
\label{phase}
\end{eqnarray}

The Equation (\ref{KE}) should be supplemented with an initial condition compatible with the requirement of electroneutrality (\ref{fe}). 
The simplest option $f_0(\vec{p})=f(\vec{p},t=t_0)=0$ corresponds to the absence of excitations at the initial moment of time $t_0$.

For the first time, KE of this type were obtained in independent works \cite{Grib:1988,Bialynicky-Birula:1991, Schmidt:1998} in the framework of standard QED. 
The specifics of the quantum field system under consideration consists in the ability to perform canonical Bogolyubov transformations in an explicit form (Section \ref{sect:tqr}).
Note that the KE (\ref{KE}) is valid in the case of an arbitrary polarization of the external field (\ref{A1}).
The integro-differential KE (\ref{KE}) can be written in the form of an equivalent system of ordinary differential equations
\begin{eqnarray}
\dot{f}=\frac{1}{2}\lambda u,\quad \dot{u}=\lambda \left( 1-2f\right) -\frac{2\varepsilon}{\hbar }
v ,\quad \dot{v}=\frac{2\varepsilon }{\hbar }u
\label{KESyst}
\end{eqnarray}
with appropriate initial conditions
\begin{eqnarray}
f_0(\vec{p})=u_0(\vec{p})=v_0(\vec{p})=0~.
\label{fuv}
\end{eqnarray}
The auxiliary functions $u(\vec{p},t)$ and $v(\vec{p},t)$ in the ODE system (\ref{KESyst}) describe vacuum polarization effects and can be written in terms of anomalous averages (\ref{DFanomal}), (\ref{DFanomal2}) 
\begin{eqnarray}\label{uvdef}
v=i\left[ f^{(+)}-f^{(-)}\right] ,
u = f^{(+)}+f^{(-)}.
\end{eqnarray}
The KE system (\ref{KESyst}) has an integral of motion
\begin{eqnarray}
(1 - 2f)^2 + u^2 +v^2 = 1,
\label{int_m}
\end{eqnarray}
compatible with the initial conditions (\ref{fuv}).
An interesting property of the KE (\ref{KE}) (or its equivalent system (\ref{KESyst})) is the preservation of the non-negativity of the distribution function $f(\vec{p},t)\geq 0$  in the entire domain of its definition.

\section{Macroscopic Averages \label{sect:ma}}

Since, due to the uncertainty relation, the description of the system in terms of time-dependent quasienergy (\ref{energy}) is conditional and acquires physical meaning only in the asymptotic region $ t\to \infty $,  the distribution function itself is rather {conventional}. This feature of the quasiparticle approach was noted in the literature (for example, \cite{Birrell:1982}). On the other hand, it is understood that under certain conditions (for example, in the case of systems located in a limited region of space, $ V<\infty $), the evolution of a macroscopic quantum field system in time-dependent external conditions can be controlled at any time in various ways (using radiation into the external region or various responses to weak external probes). Such, for example, are experiments on graphene samples in the optical excitation range \cite{Gierz:2013, Baudisch:2018} or planned experiments to detect an $e^-e^+$   plasma in the focal spot of computer-propagating powerful laser beams \cite{Kawanaka:2016, Shen:2018, Tanaka:2020}.

In such a situation, macroscopic averages obtained by taking the expectation value with $|in\rangle$ states and integrating the dynamic characteristics of quasiparticles over the momentum space, statistically weighted with the distribution function $ f(\vec{p},t) $ playing the role of time-dependent functions mediating between the dynamic characteristics of quasiparticles and physical observables.

The density of excited electron-hole ($ e-h $) pairs is the simplest quantity of this type.
\begin{eqnarray}
n(t)=\frac{N_f}{L^2}\sum_{\vec{p}}f(\vec{p},t),
\label{nt}
\end{eqnarray}
where $ N_f=N_S N_D=4 $ is the total number of flavors in the model under consideration.

From the full Hamiltonian of the fermionic subsystem (\ref{Hferm}) follows the expression for the total energy density of the electron - hole subsystem, which consists of the sum of the quasiparticle and polarization parts
\begin{eqnarray}
E_f(t)=E_{eh}(t)+E_{pol}(t),
\label{Eft}
\end{eqnarray}
where
\begin{eqnarray}  
\label{Eeht}
E_{eh}(t)&=&2\frac{N_f}{L^2}\sum_{\vec{p}}\varepsilon(\vec{p},t)f(\vec{p},t), \\
E_{pol}(t)&=&-\frac{1}{2}\frac{N_f}{L^2}\sum_{\vec{p}}\hbar\lambda(\vec{p},t)v(\vec{p},t),  %\notag
\label{Epolt}
\end{eqnarray}
where $ v(\vec{p},t) $ is the vacuum polarization function (\ref{uvdef}).

The current density can also be represented as the sum of the conduction and polarization currents
\begin{eqnarray}
J_k(t)=J_k^{cond}(t)+J_k^{pol}(t),
\label{J1}
\end{eqnarray}
where
\begin{eqnarray}
J_k(t)=\frac{1}{L^2}\int d^2xj_k(\vec{x},t).
\label{J2}
\end{eqnarray}

To write these currents in terms of the functions $ f,u,v $, we use the formula (\ref{CurDens}) and perform a unitary transition to the quasiparticle representation there (Section 3), 
\begin{eqnarray}
J_k(t)=e{\mathrm v}_F\frac{N_f}{L^2} \sum_{\vec{p}}\Phi^\dagger \mathcal{U}^\dagger \sigma^k\mathcal{U}\Phi.
\label{J3}
\end{eqnarray}

Using formulas (\ref{unitar}), (\ref{components}) and the definition (\ref{uvdef}) here, we obtain the expression for the conduction current density as a result
\begin{eqnarray}
J_k^{cond}(t)=2e\frac{N_f}{L^2} \sum_{\vec{p}}v^k_q(\vec{p},t)f(\vec{p},t)
\label{J4}
\end{eqnarray}
where
\begin{eqnarray}
v^k_q(\vec{p},t)=\frac{\partial\varepsilon(\vec{p},t)}{\partial P^k}
=\frac{{\mathrm v}_F^2P^k}{\varepsilon(\vec{p},t)}
\label{vkq}
\end{eqnarray}
is the propagation velocity of the quasiparticle excitation. In the derivation of formula (\ref{J4}), the vacuum unit was omitted  $ 2f-1 \to 2f $. The polarization current density is
\begin{eqnarray}
J_k^{pol}(t)=-e\frac{N_f}{L^2} \sum_{\vec{p}}\tilde{v}^k_q(\vec{p},t)u(\vec{p},t),
\label{J5}
\end{eqnarray}
where the vector of ``conjugate velocity'' $ \tilde{v}^k_q(\vec{p},t) $ is determined through the components of the vector $ v^k_q(\vec{p},t) $
\begin{eqnarray}
\tilde{v}^k_q=(v^2_q~,-v^1_q).
\label{vkq2}
\end{eqnarray}

Using this definition, one can write the amplitude (\ref{lambda}) in the following form:
\begin{eqnarray}
\lambda=\frac{e}{\varepsilon}\vec{E}\vec{\tilde{v}}_q.
\label{lambdaEv}
\end{eqnarray}

Thus, the functions $ u(\vec{p},t) $ and $ v(\vec{p},t) $ determine the energy and current due to the polarization of the medium.
We now obtain the law of conservation of energy in the quasiparticle subsystem. Differentiating in time the density of quasiparticle energy (\ref{Eeht}) and using the formulas (\ref{J4})--(\ref{vkq2}) for the current density and relation (\ref{lambdaEv}), we obtain
\begin{eqnarray}
\dot{E}_{eh}(t)=\vec{E}(t)\vec{J}(t),
\label{Eeht2}
\end{eqnarray}
where  $ \vec{J}(t) $ is the total current density (\ref{J1}).
This relation allows a different formulation. We write the Maxwell equation for the internal plasma field
\begin{eqnarray}
\dot{\vec{E}}_{in}(t)=-\vec{J}(t)
\label{EintJ}
\end{eqnarray}
and write in Equation (\ref{Eeht2}) the total electric field as a sum
\begin{eqnarray}
\vec{E}(t)=\vec{E}_{in}(t)+\vec{E}_{ex}(t).
\label{totalE}
\end{eqnarray}

As a result, we obtain
\begin{eqnarray}
\frac{d}{dt} \left[ \vec{E}_{eh}(t)+\frac{1}{2}\vec{E}^2_{in}(t)\right] =\vec{J}(t)\vec{E}_{ex}.
\label{dEJ}
\end{eqnarray}

Similarly, relation (\ref{Epolt}) implies the relation describing the energy balance of vacuum polarization
\begin{eqnarray}
\dot{E}_{pol}(t)=-\vec{E}(t)\vec{J}^{pol}(t)-\frac{\hbar}{2}\frac{N_f}{L^2}\sum_{\vec{p}}\dot{\lambda}(\vec{p},t)v(\vec{p},t).
\label{Evac}
\end{eqnarray}

In the next sections the electrodynamics of graphene as well as some features of the inner plasma field will be considered. Section \ref{sect:cf} details the case of a constant electric field, and Section  \ref{sect:hfp} the case of an alternating field.

\section{Constant Field \label{sect:cf}}

We begin the demonstration of the opportunities of the developed approach with the case of a constant external electric field:
\begin{eqnarray}
\vec{E}(t)=const .
\label{Econst}
\end{eqnarray}

Constant is both the absolute value of the field strength and its direction.
Further, the first axis of the used two-dimensional coordinate system is associated with this direction.
In this case, it is convenient to compare with the results of using the Landauer approximation (or the Landauer-Datta-Lundstrom (LDL) model) \cite{Landauer:1957, Landauer:1970, Landauer:1996} for calculating the characteristics of the current arising in graphene samples with the results of the presented approach.
Some nontriviality is that quite different methods and different spaces of description are used here: the considered kinetic approach is formulated in the $t$-representation, while the LDL method based on the WKB approximation uses the $x$-representation. Both approaches are equivalent in the case of a constant electric field.

A constant field does not have any characteristic time scales. 
Those can only be sought in the characteristics of the material itself or of the simulated sample. For a material (graphene), the natural unit of the time scale is the ratio of the lattice constant to the Fermi velocity 
$ a/{\mathrm v}_F\approx 2.46\times10^{-16}$ s. 
It~makes no sense to consider processes on a smaller time scale.
%Further, this value will be used as a unit of the natural time scale.
The upper boundary of the time scale is determined by the condition of applicability of the assumption of spatial homogeneity of the system. It depends on the characteristic size of the sample and can be determined by the ratio 
$ L/{\mathrm v}_F $.

The system of Equation (\ref{KESyst}) allows only a numerical investigation. For each point of the momentum space $\left\lbrace p_1,p_2 \right\rbrace $, it is solved independently. 
The region in which the distribution function is localized and the required density of its coverage by the computational grid at the nodes of which solutions for (\ref{KESyst}) will be sought is determined by the necessary accuracy in calculating the integral characteristics. 
This is realized by a sequential iterative procedure with stepwise control of the accuracy of the results obtained.

\begin{figure}[!h]
	\includegraphics[width=0.50\textwidth]{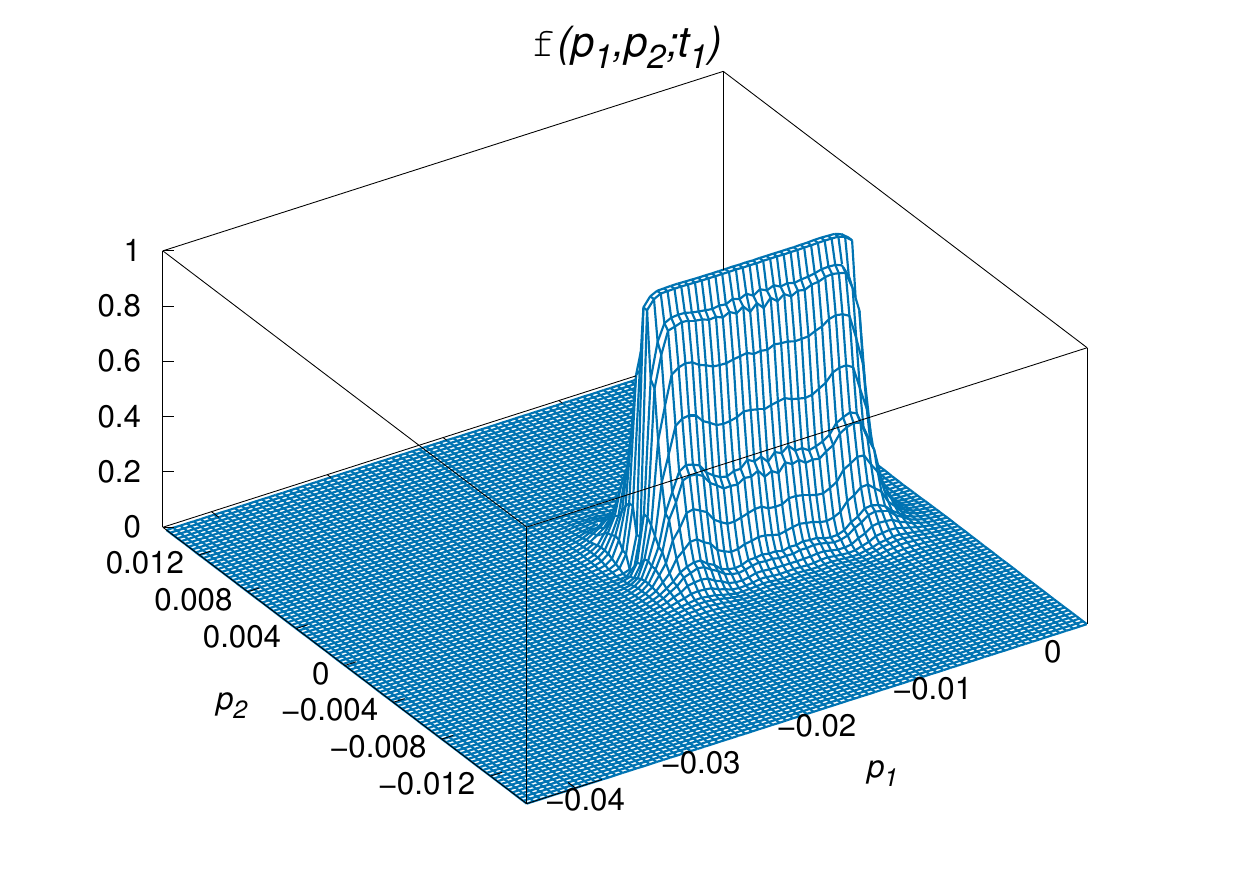} \hfill
	\includegraphics[width=0.50\textwidth]{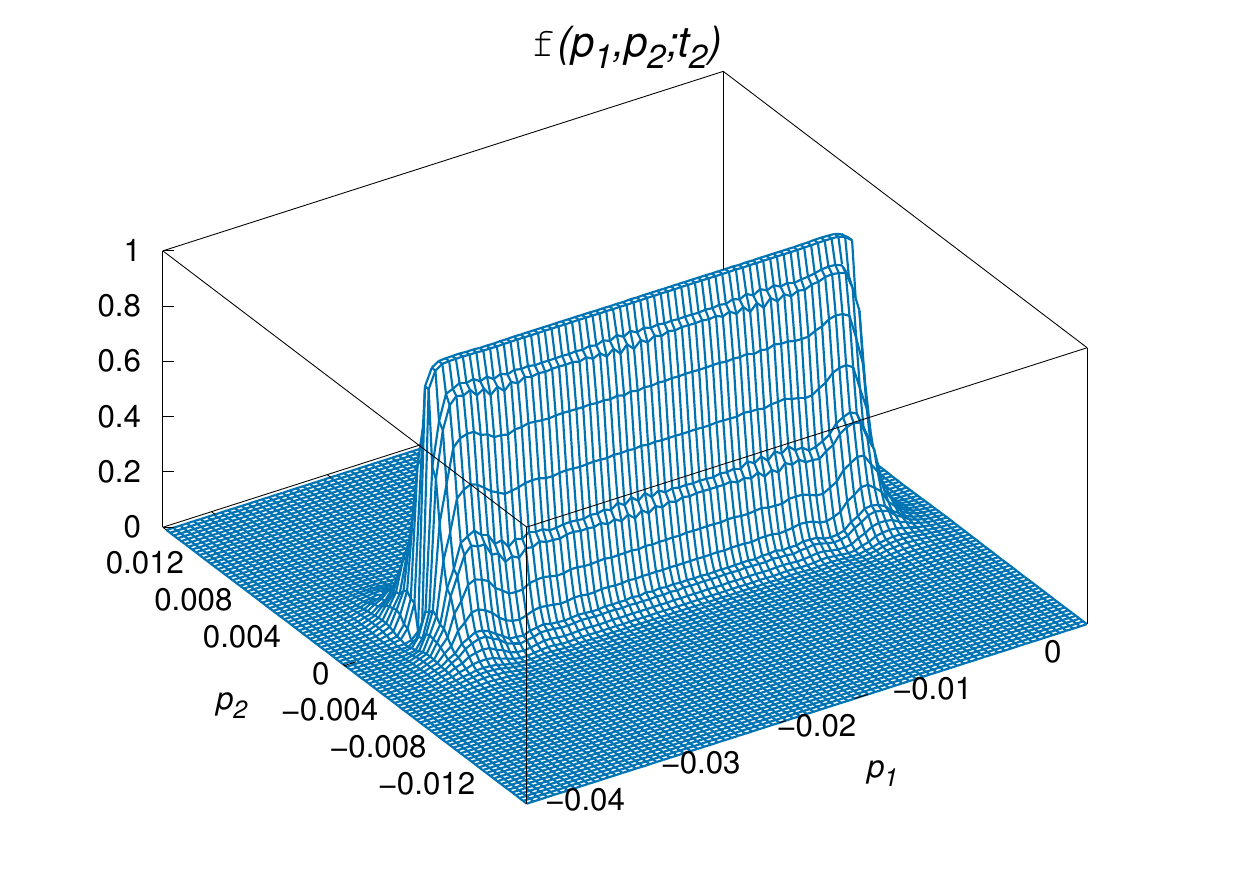} 
	\caption{The distribution function in the $2D$ momentum space for two consecutive points of time at the field strength $0.1$ V/$\mu$m. {\bf  Left panel}: $t_1=0.5\times10^{-12}$ s, {\bf Right panel}: $t_2=1.0\times10^{-12}$ s.
		\label{fig:1}}
\end{figure}

Figure~\ref{fig:1} shows the form of the distribution function for two consecutive points of time at a field strength of 
$0.1$ V/$\mu$m. 
The ``natural'' value $\hbar/a$ is used hereinafter as a unit value for the $p_1,p_2$. 
Such~kind of accumulative behavior of the distribution function is in agreement with the results of~\cite{Fillion-Gourdeau:2015} and \cite{Li:2018}. 
The latter work based on the Greens function method takes into account the dissipative mechanism of inelastic scattering of optical phonons.

\begin{figure}[!h]
	\includegraphics[width=0.5\textwidth]{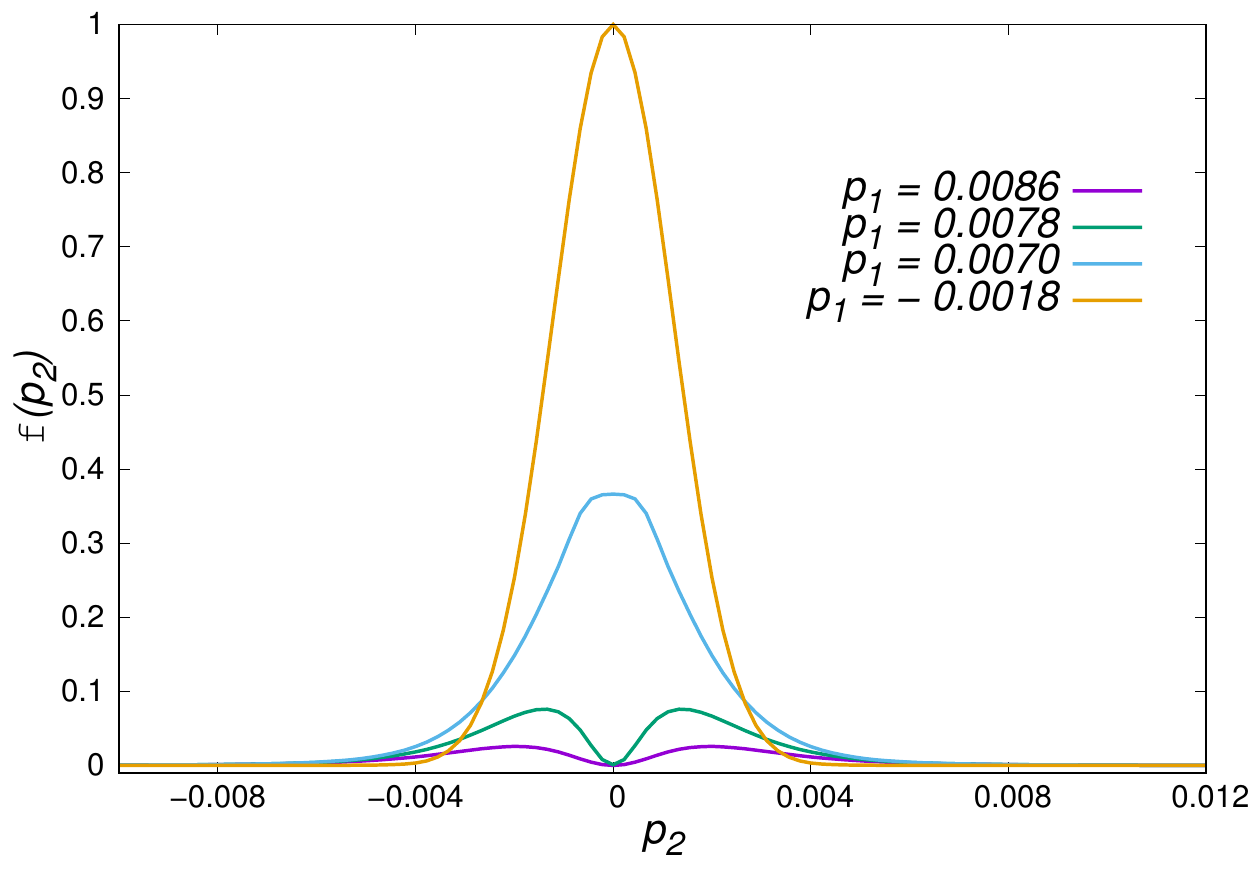}\hfill
	\includegraphics[width=0.5\textwidth]{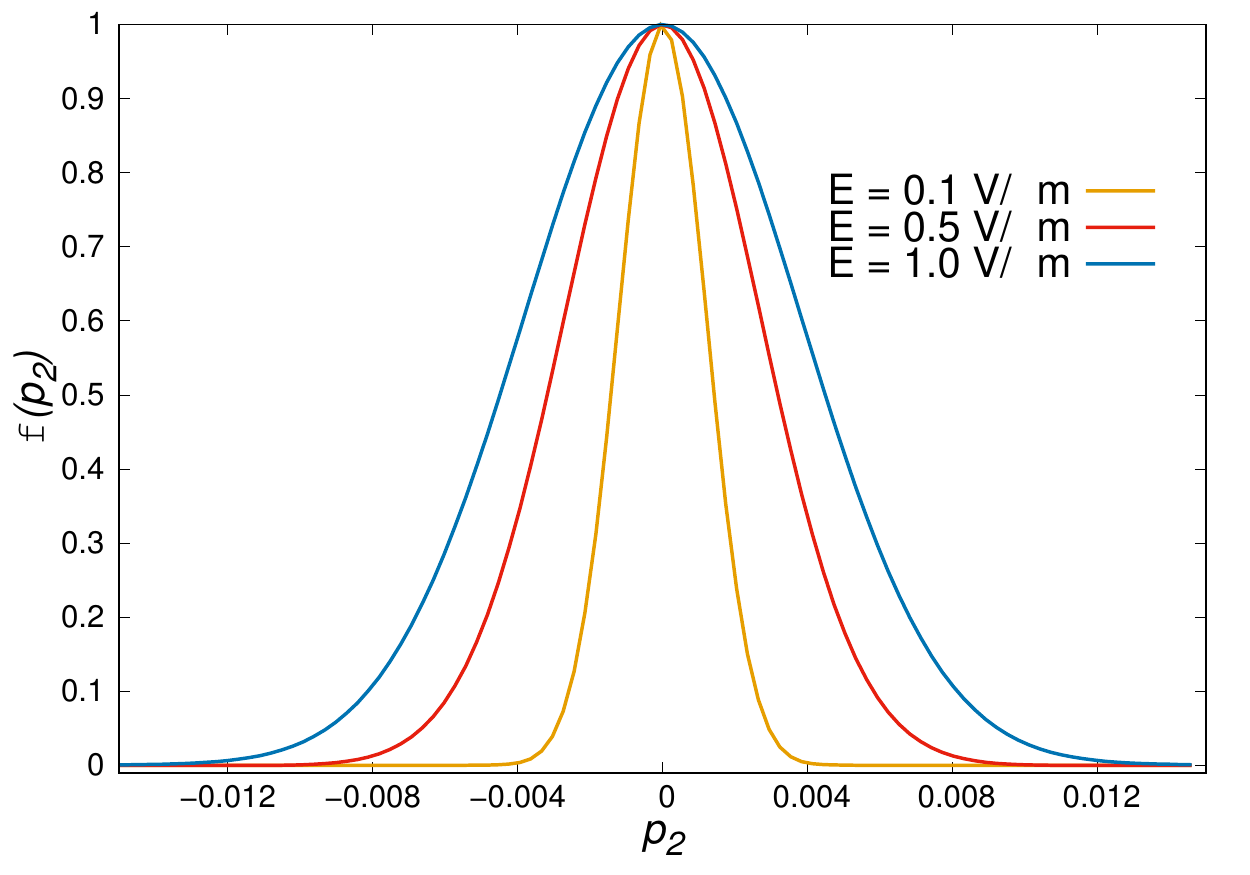}
	\caption{{\bf  Left panel}: The section of the distribution function $f(p_1,p_2)$ for several values of $p_1$ at the field strength $0.1$ V/$\mu$m; 
	{\bf Right panel}: The shape of the section $f(p_1,p_2)$ for $p_1=-0.0018$ for different field strength $E$.
		\label{fig:2}}
\end{figure}

The distribution function of the excitation on the initial stage is formed and reaches the maximum value $f_{\rm max}=1$ very rapidly. 
Figure~\ref{fig:2}, left panel, shows the transversal sections of this distribution for different values $p_1$ at early time of the evolution, corresponding to the right edges of the distribution functions in Figure~\ref{fig:1}. 
In the case of the ballistic regime the following accumulation of population is a result of the increase of the longitudinal momentum $p_1$ in the direction of the acting electric field at invariable Gauss-like distribution on the transversal momentum $p_2$. The halfwidth of the $p_2$-distribution is defined by the field strength $E$ (Figure~\ref{fig:2}). 
As a result, the number density of carriers grows proportionally to the action time $T$ of the field,  $n(T)\propto T$,  as in the case of standard QED (see, e.g., \cite{Grib:1988}).

In the LDL approach, the problem of calculating the current density through the sample of a material of finite width $L\gg a = 0.246$ nm limited by two parallel electrodes is solved by calculating the transmission probability $T(\varepsilon,p_2,{\rm U})$  in the presence of a given potential difference ${\rm U}$ {(for dispersion of graphene $\varepsilon = {\mathrm v}_{F}\sqrt{(p_{1})^{2}+(p_{2})^{2}}$)} \cite{Vandecasteele:2010}:
\begin{eqnarray}
j_{LDL}=\frac{4e}{(2\pi \hbar)^2}\int dp_2 \int_{\varepsilon_F-e{\rm U}}^{\epsilon_F} d\varepsilon T(\varepsilon,p_2,{\rm U}).
\label{jLDL}
\end{eqnarray}

Therefore, the momentum component $p_2$ does not change its value during tunneling. In spite of the fact that graphene is a gapless semiconductor, the presence of finite conserved values $p_2$ leads to the appearance of an energy gap with the width $\Delta = 2{\mathrm v}_Fp_2$. In the considered case of a vacuum the initial state temperature and chemical potential are equal to zero, the Fermi energy $\varepsilon_F=0$ and there are no free carriers in the interelectrode space. Under these conditions the process of  carrier transmission can proceed only by Zener-Klein tunneling. The probability of this tunneling in the WKB approximation is (taking into account the relationship of $\varepsilon$ and $p_2$):
\begin{eqnarray}
T(\varepsilon,p_2,{\rm U})=T_{ZK}=\exp \left( -\frac{\pi p^2_2 {\mathrm v}_FL}{e \hbar {\rm U}}\right). 
\label{TEpv}
\end{eqnarray}

In this case from (\ref{jLDL}) and (\ref{TEpv}) one obtains \cite{Vandecasteele:2010}:
\begin{eqnarray}
j_{LDL}&=&2\frac{{\rm U}e^2}{\pi^3 \hbar L}\sqrt{\frac{\pi^2e{\rm U}L}{4\hbar {\mathrm v}_F}}\times\left(\mathrm{erf} \left[  \sqrt{\frac{\pi e{\rm U}L}{4\hbar {\mathrm v}_F}}\right] \right.\nonumber\\
&&\left.+ \exp \left[ -\frac{\pi e{\rm U}L}{4\hbar {\mathrm v}_F} \right] -1   \right). 
\label{jLDL2}
\end{eqnarray}

First of all, let us note that the distribution (\ref{TEpv}) of carriers over the transverse momentum $p_2$ reproduces the result (Figure~\ref{fig:2}, right panel) obtained in the kinetic approach to the permille accuracy, i.e., within the thickness the lines.

The basic problem is that the conduction current density (\ref{J4}) 
%% (\ref{CurDens})%% 
in the absence of dissipation and spatial boundaries depends on the time after switching on the field and increases continuously, which, obviously, is not observed in the experiment, where each value of the potential difference corresponds to its steady-state current density. 
So, under the conditions of the real experiment \cite{Vandecasteele:2010} it is necessary to take into account the presence of the electrodes, which limit the lifetime of the carriers. 
From this point of view the process of carrier generation must permanently continue throughout the entire measurement process and be uniform in the area of the sample. 
Knowing the strong anisotropy of the carrier spectrum, it is possible to assume in the first approximation that all of them move towards the electrode of the corresponding polarity with the velocity ${\mathrm v}_F$. 
In this case, the average lifetime of carriers is $\tau = L/2{\mathrm v}_F$ $0.5\times10^{-12} s$ for $L=1.0~\mu$m. 
At the end of this time after switching on the field, the rate of generation of carriers becomes equal to the rate of their escape through the electrodes. 
The~steady-state current values will be constant and can be calculated in both approaches (Table \ref{tab1}). 

\begin{table}[!h]
	\centering
	\caption{Comparison of calculated current density values for a sample with $L=1.0\, \mu$m in the range of potential differences from $0.1$ V to $0.5$ V. \label{tab1}}
	\begin{tabular}{ccc}
		\toprule
		{U[V]}& {\emph{j}} {in WKB Approach} & {\emph{j}} {in Kinetic Approach}\\
		{(\emph{E}= U/\emph{L} [V/}{$\mu$}{m])} & {[}{$\mu$}{A/nm]} & {[}{$\mu$}{A/nm]} \\
%		\midrule
		0.1 & 0.02904 & 0.02964\\
		%\hline
		0.2 & 0.028344 & 0.028457\\
		%\hline
		0.3 & 0.15435 & 0.15557\\
		%\hline
		0.4 & 0.23861 & 0.24003\\
		%\hline
		0.5 & 0.33440 & 0.33601\\
%		\bottomrule
	\end{tabular}
	
\end{table}

The results of these calculations coincide with high degree of accuracy in the considered range of parameters. Indirectly this shows a good coincidence with experiment \cite{Vandecasteele:2010}. The given list of values strictly corresponds to the law $I\sim {\rm U}^{3/2}\sim E^{3/2}$.
This dependence is corroborated by other sources~\cite{Gavrilov:2012jk, Klimchitskaya:2013fpa}.

\begin{figure} [!h]
	\begin{center}
	\includegraphics[width=0.48\textwidth]{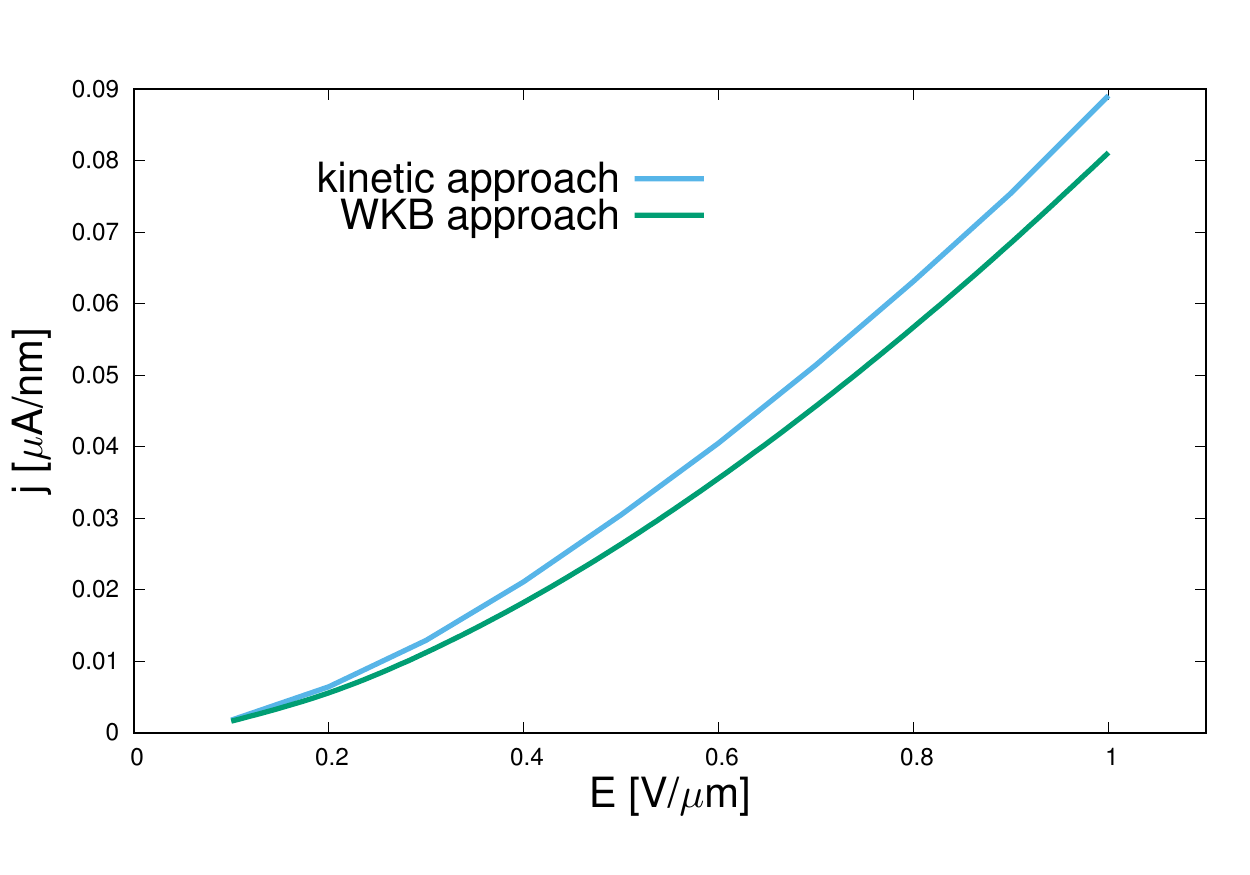}
	\end{center}
	\caption{
		Comparison of calculated {conduction} current density for a sample with $ L=100$ nm in the range of electrical field strengths $E$ from 0.1~V/$\mu$m to 1.0~V/$\mu$m (potential differences from 0.01~V to~0.1~V).
		\label{fig:3}}
\end{figure}

The kinetic method presented here can be a valid outside the framework of the applicability of the WKB approach, for example, in alternating electric fields in the region of sufficiently high frequencies. An indirect confirmation of this is the difference in the results for a thinner (tenfold) sample with a correspondingly shorter carrier lifetime (Figure~\ref{fig:3}). In this case the difference reaches almost $10\%$. 
Both, the kinetic and the WKB approach, predict superlinear growth of the current density with almost coinciding exponents $\sim 1.70$.

\begin{figure}[!h]
\begin{center}
	\includegraphics[width=0.48\textwidth]{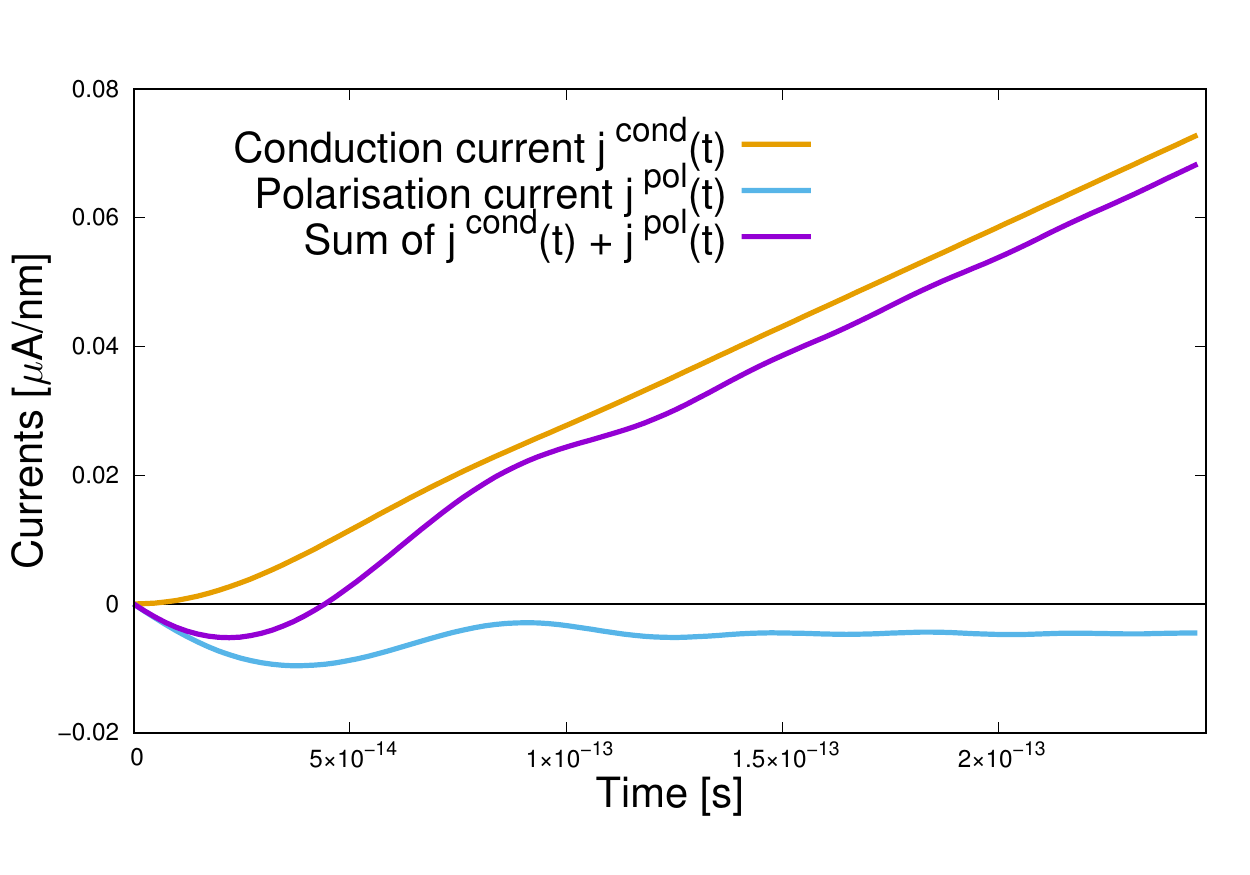}
\end{center}
	\caption{Conduction current (upper curve), polarization current (bottom curve) and their sum as a function of time for $t \ll L/{{\mathrm v}_F} $.
	\label{fig:4}}
\end{figure}

Above, we limited ourselves to considering only the conduction current in the situation of a {stationary process}, when the number of carriers born compensates for their departure from the sample. 
In the kinetic approach, the behavior of the conduction current (\ref{J4}) and the polarization current (\ref{J5}) can be reproduced at times $t \ll L/{{\mathrm v}_F} $, when carrier losses at the sample boundaries can be neglected. 
The change in the surface density of the conduction current, polarization current and their sum after turning on the constant field of $ 0.3$ V/$\mu$m directed along the $x$-axis is shown in Figure~\ref{fig:4}. 

The behavior of the polarization current is fundamentally different from conduction current behavior and has the form of damped oscillations asymptotically reaching a constant negative value.
Nevertheless, the contribution of the conduction current becomes dominant quite quickly.

\section{Short High Frequency Pulses \label{sect:hfp}}

Next, we consider the field model of pulses with a cyclic carrier frequency $\omega$ and a Gaussian envelope with duration parameter  $\tau$, 
\begin{equation}
E(t)=E_a \cos (\omega t+\varphi)e^{-t^2/2\tau^2} . 
\label{Eta}
\end{equation}

The phase shift $ \varphi $ sets the position of the absolute maximum of the field relative to the maximum value of the envelope (which will be associated with the time $t=0$). The spatial direction of the field is constant. Below we stay within the limitations of the low-energy model and consider the region of rather small field strengths and frequencies. In the case of strong fields it is necessary to use some generalizations of this model. For example, a corresponding generalization of the kinetic theory was developed in the work \cite{Smolyansky:2019} for the tight binding model of the nearest neighbour interaction \cite{Kao:2010, Castro:2009, Gusynin:2007}.

\begin{figure}[!h]
\begin{center}
	\includegraphics[width=0.48\textwidth]{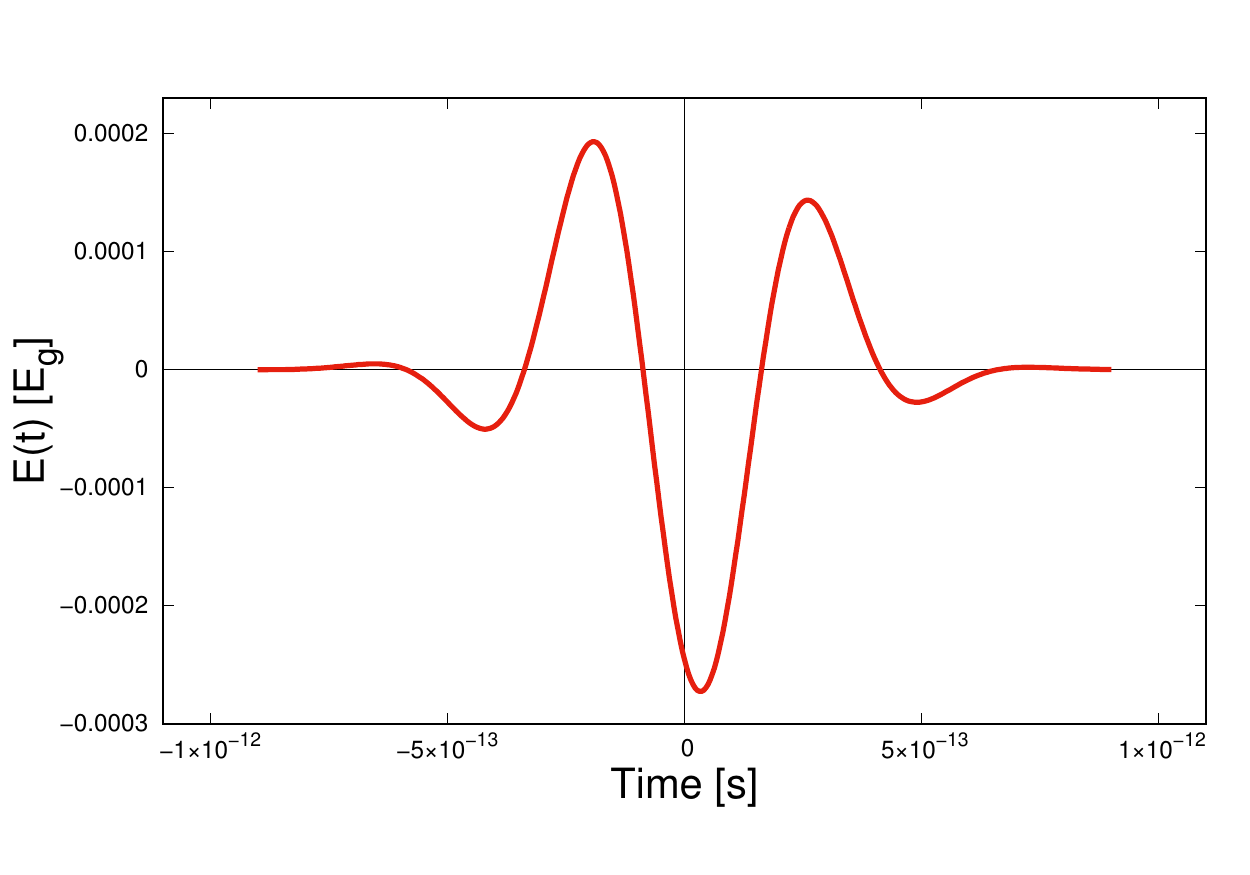}
\end{center}
	\caption{{Time dependence of the field strength for the Gaussian} envelope harmonic pulse (\ref{Eta}) for $E_a = 3.0$ V/$\mu$m, $\omega=2\pi \times2$ THz, $\varphi=0.85~\pi$ and $\tau=3/\omega \approx 2.4\times10^{-13}$ s. 
		\label{fig:6}}
\end{figure}

We choose the characteristics of the field so that they correspond to the parameters of the experiments described in \cite{Bowlan:2014}. These are very short pulses with $\omega=2\pi \times2$ THz carrier frequency and duration parameter $\tau=3/\omega \approx 2.4\times10^{-13}$ s, which almost coincides with the constant-field action time considered above. But since in this case the field turns on and off relatively slowly, to accurately reproduce the behavior of the model, we will consider a several times longer interval. The electric field amplitude is $E_a = 3.0$ V/$\mu$m, which is an order of magnitude greater than that considered above for a constant field. To accurately reproduce the parameters, we used a nonzero value of the carrier phase shift 
$\varphi=0.85~\pi$. 
An explicit form of the dependence of the field on time is shown in Figure~\ref{fig:6}. 
The field is formed by a linearly polarized electromagnetic wave and is assumed to be directed along the $x$-axis.
%
%Here the units are the same as in Figure~\ref{fig:4}. 
The field strengths are expressed as $ E(t)/E_g $, where  $E_g=\hbar {\mathrm v}_F/e a^2 = 1.088 \times 10^4$ V/$\mu$m.

The distribution function formed under the action of such a field through $6.0\times 10^{-13}$ s after passing the maximum (maximum envelope, since in this case $\varphi \neq 0$), is presented in the following Figure~\ref{fig:7}: 

\begin{figure} [!h]
\begin{center}
	\includegraphics[width=0.48\textwidth]{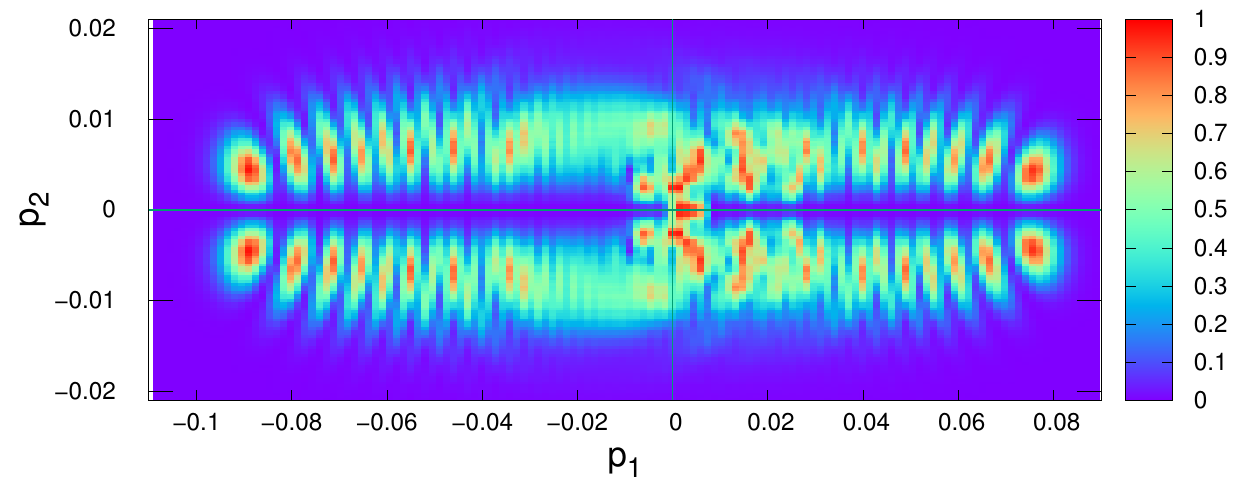}
\end{center}
	\caption{Distribution function $f(p_1,p_2)$ formed under the action of the field (\ref{Eta}) shown in Figure~\ref{fig:6} at $6.0\times 10^{-13}$ s after passing the maximum of the field strength.
		\label{fig:7}}
\end{figure}

The distribution function is localized in the vicinity of the Dirac point. The distribution of populated states is much more complicated than in the case of a constant field. Thus reproduction of integral characteristics in this case may be more time consuming. The figure is based on the results of solving kinetic equations on a maximally simplified $222 \times 46$ grid. 
The carrier density (\ref{nt}) evolution reproduced from these data for a time interval from $-6.0\times 10^{-13}$ s to 
$ 6.0 \times 10^{-13}$ s is shown in the following Figure~\ref{fig:8}:

\begin{figure}[!h]
\begin{center}
	\includegraphics[width=0.48\textwidth]{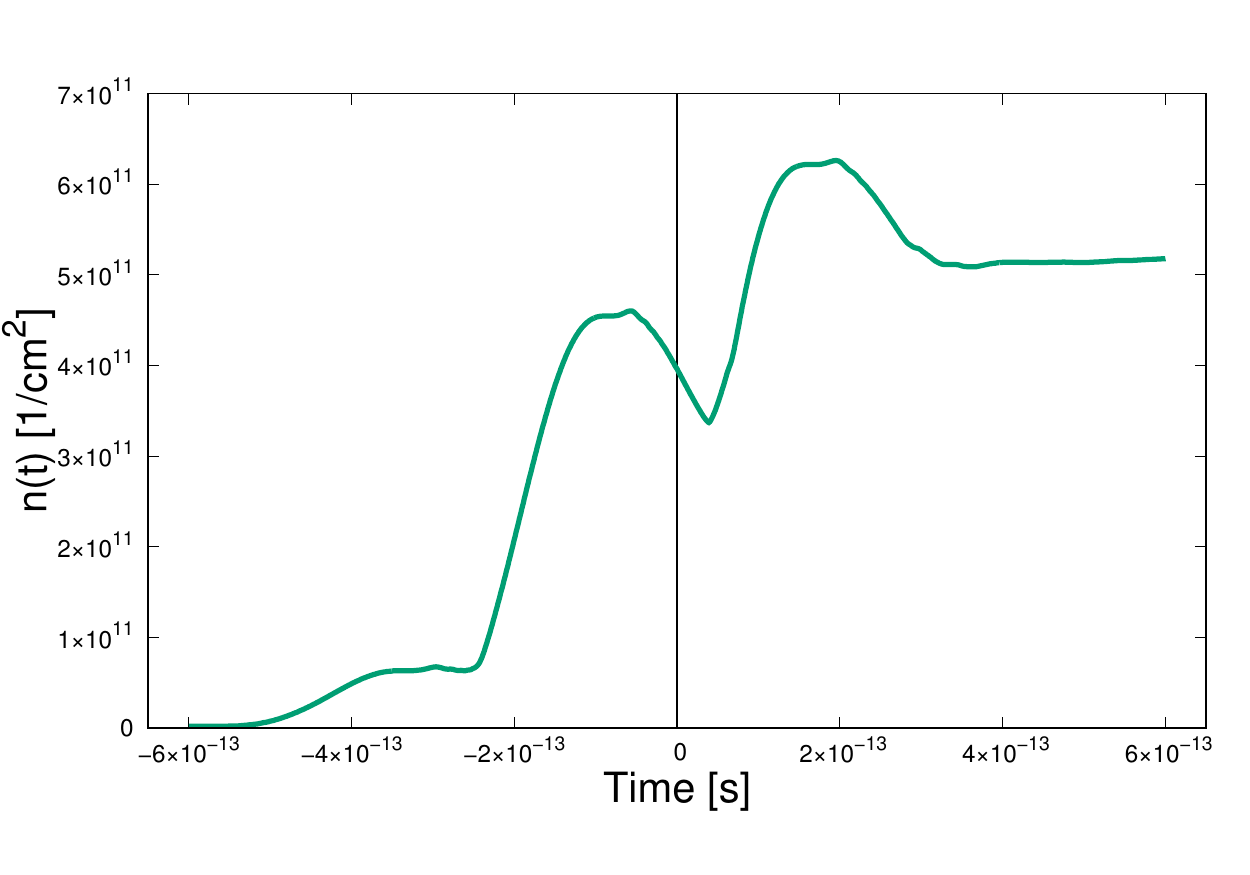}
\end{center}
	\caption{{The carrier density evolution for the field of }Figure~\ref{fig:6} for a time interval from $-6.0\times 10^{-13}$ s to $ 6.0 \times 10^{-13}$ s.
		\label{fig:8}}
\end{figure}

Starting from zero, the carrier density is quite difficult to evolve and reaches a final constant value, which is the residual carrier density. A short field pulse generates several carriers, which, in the absence of dissipation, continue to exist even after the field is turned off.

\begin{figure}[!h]
\begin{center}
	\includegraphics[width=0.48\textwidth]{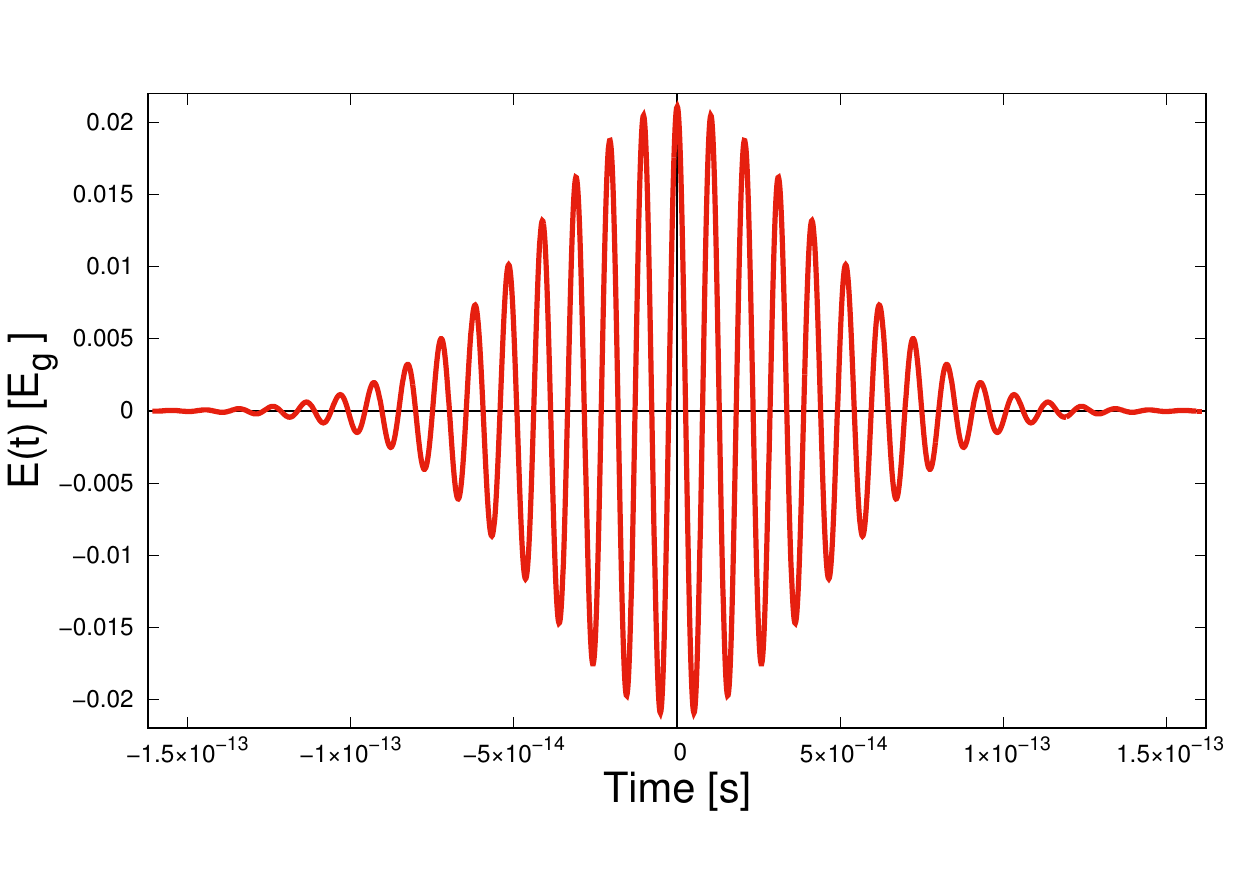}
\end{center}
	\caption{Time dependence of the field strength for the Gaussian envelope harmonic pulse (\ref{Eta}) for 
	$E_a = 2.2\times 10^{2}$ V/$\mu$m, $\omega = 2\pi \times 96.7$ THz, $\varphi=0$ and 
	$\tau \approx 26/\omega \approx 4.28\times 10^{-14}$ s.
		\label{fig:9}}
\end{figure}

We consider one more variant of a field of the form (\ref{Eta}). 
This will be a relatively long infrared pulse, similar in parameters to those considered in \cite{Baudisch:2018}. 
We define its cyclic frequency as $\omega = 2\pi \times 96.7$~THz, which corresponds to a wavelength of $3.1\,\mu$m and a photon energy of $0.4$ eV. 
The duration parameter is $\tau \approx 26/\omega \approx 4.28\times 10^{-14}$ s. 
The phase shift was chosen equal to zero. 
The maximum electric field strength was determined based on the declared energy flux density in the focal spot of $7$ GW/cm$^2$ and reaches $\approx 2.2\times 10^{2}$ V/$\mu$m and exceeds by almost two orders of magnitude the corresponding value from the previous example. 
%In units of $ E(t)/E_g$ and the natural time scale, 
The impulse with these characteristics has the form shown in Figure~\ref{fig:9}.
%As in the previous case, the field is assumed to be directed along the $x$-axis.

\begin{figure} [!h]
 \begin{center}
 	\includegraphics[width=0.48\textwidth]{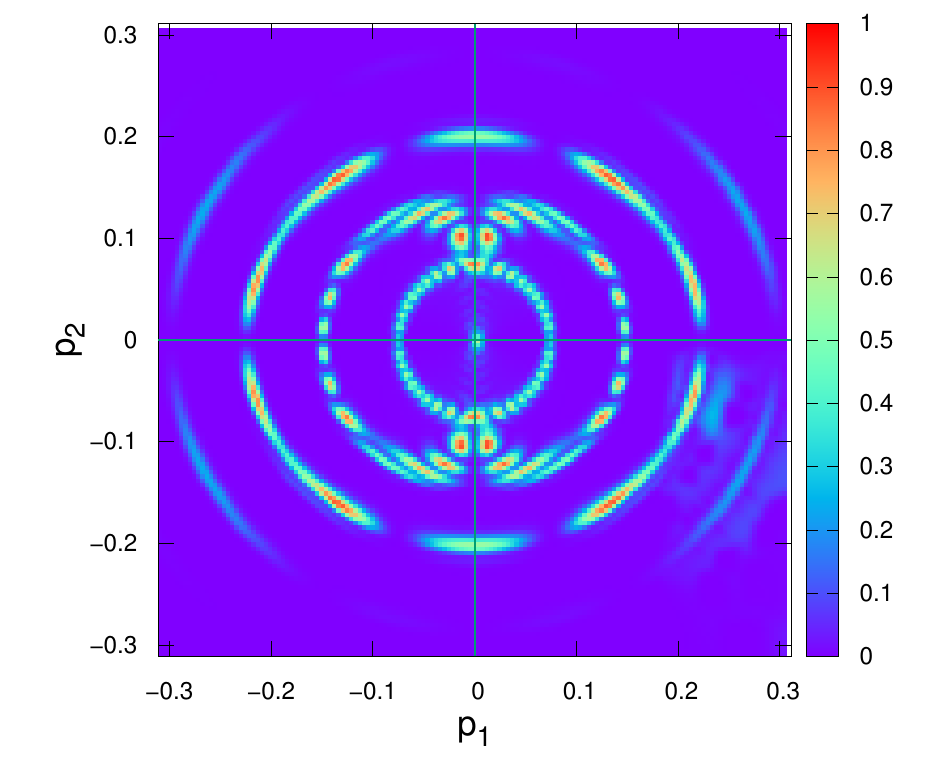}
\end{center}
 	\caption{Momentum space distribution function $f(p_1,p_2)$ at the final stage of the action of the external~field.
 		\label{fig:10}}
 \end{figure}

The form of the distribution function at the final stage of the action of the external field is shown in Figure ~\ref{fig:10}.

\begin{figure} [!h]
\begin{center}
	\includegraphics[width=0.48\textwidth]{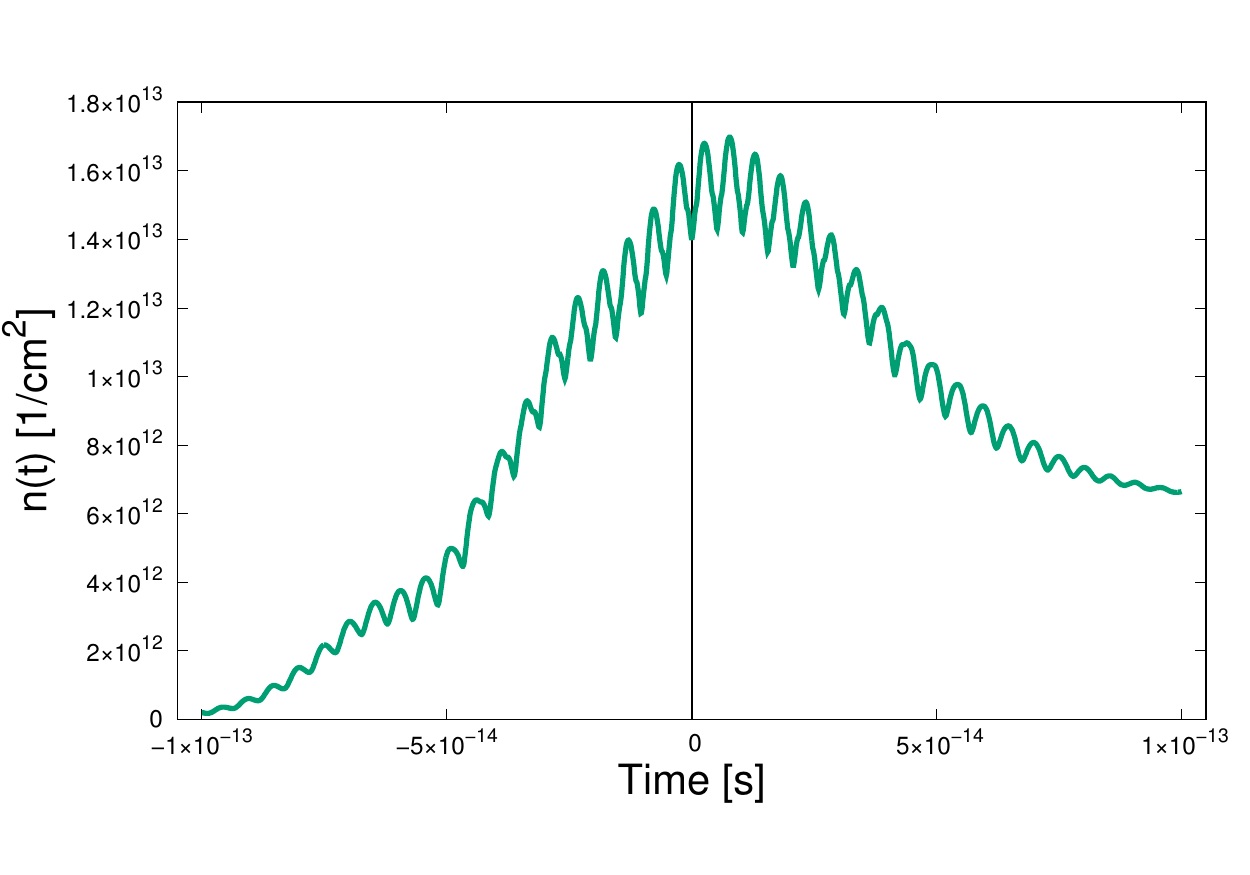}
\end{center}
\vspace{-6pt}
	\caption{{Time dependence of the carrier density for the infrared pulse of Figure} \ref{fig:9}.
		\label{fig:11}}
\end{figure}

The region in which non-zero values of the distribution function are observed in this case is approximately $ -0.30 \leq p_1 \leq 0.30, -0.25 \leq p_2 \leq 0.25 $. The distribution of occupied states has a characteristic ring structure. Due to the great computational complexity in this case, the results are presented for a $64 \times 64$ grid. It allows getting a general idea about the structure of the distribution function. The evolution of the carrier density is shown in Figure \ref{fig:11}:

At the stage of increasing field amplitude, an increase in the density of carriers with small oscillations is observed. At the stage of turning off the field, a drop to a certain residual value takes~place.

\begin{figure} [!h]
\begin{center}
	\includegraphics[width=0.48\textwidth]{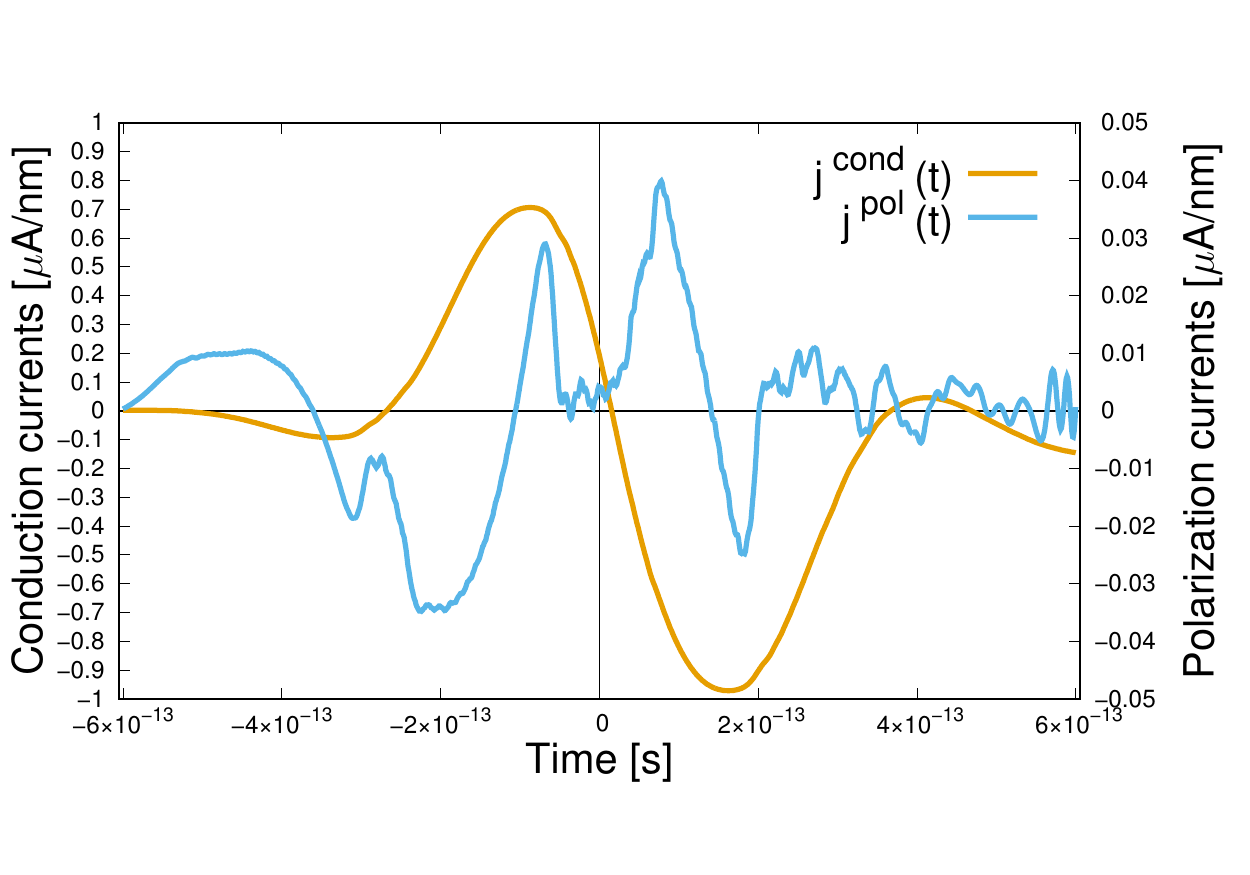}
\end{center}
\vspace{-6pt}
	\caption{Time dependence of the conduction and polarization currents for a pulsed alternating field with a cyclic frequency $ \omega = 2\pi \times 2$ THz.
		\label{fig:12}}
\end{figure}

Conduction and polarization currents for a pulsed alternating field with a cyclic frequency 
$ \omega = 2\pi \times 2$ THz (Figure \ref{fig:6}) are shown in Figure \ref{fig:12}.

For the conduction current, one can note some qualitative analogy with the time dependence of the acting field.The nature of the dependence on the time of the polarization current is more complicated.

\begin{figure}[!h]
\begin{center}
	\includegraphics[width=0.48\textwidth]{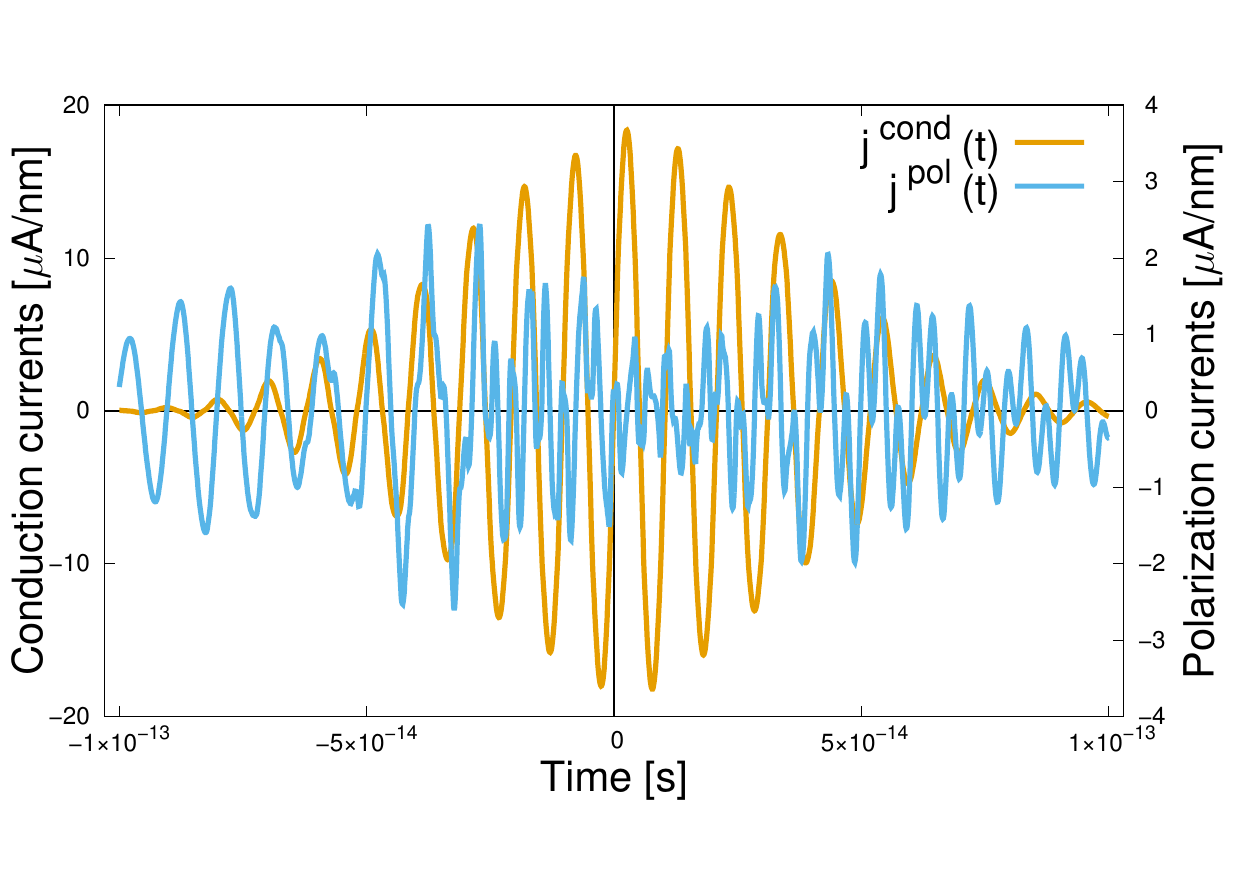}
\end{center}\vspace{-6pt}
	\caption{Time dependence of the conduction and polarization currents for a pulsed alternating field with a cyclic frequency $ \omega = 2\pi \times 96.7$ THz. 
	\label{fig:14}}
\end{figure}

For an infrared pulse, the dependence of currents on time are shown in Figure \ref{fig:14}.

The conduction current demonstrates a very good reproduction of the general nature of the acting field: it is alternating, the period of polarity change obviously coincides with the period of the carrier of the simulated pulse. The amplitude of the conduction current increases and decreases along with an increase and decrease in the amplitude of the acting field. Obviously, there is a phase shift between it and the external field, because at the time $ t=0 $, when the field reaches its maximum value (and this is the absolute maximum), the conductivity current is close to zero. The behavior of the polarization current is fundamentally more complex. Over the entire time interval presented, in which the conduction current has time to increase by several of orders of magnitude and then also to decrease its amplitude, the polarization current does not show obvious signs of a regular change in its amplitude, but only irregular changes in the current value. In the initial section of the time interval presented, one can try to associate its behavior with the current values of the external field, but then a complex interference pattern is observed, which is characteristic for superposition of oscillations with different frequencies. We can draw an analogy with the slow relaxation of the polarization current already after the completion of the action of a short unipolar pulse, which was noted in \cite{Smolyansky:2017}.

% and ~\ref{fig:13}.

%has the following form:

\section{Plasma Field \label{sect:pf}}

The appearance of currents, by virtue of Maxwell's Equation (\ref{EintJ}), leads to the generation of a plasma field, which will add to the external field and in the general case will affect the process of further evolution of the system. 
We use the well-known solution to the problem of the field of a time-dependent current on an infinite plane (see, for example, Ref. \cite{Abbott:1984}): 
\begin{equation}
E(t,z)=-\frac{\mu_0 c}{2} J\left( t-\frac{z}{c} \right) . 
\label{Etdz}
\end{equation}
Here $\mu_0$ is the magnetic susceptibility of vacuum, $c$ is the speed of light and $z$ is the distance of the observation point from the graphene plane.
It follows from this expression that in the case under consideration, the plasma field is determined 
by the current density.
From the expression (\ref{Etdz}) it follows that the alternating current will generate radiation that carries information about its characteristics.
When considering a real sample with a characteristic finite size $L$, all of the above will be true only for a distance $z$ from its surface that satisfies the condition $z\ll L$. 

One of the reasons for the continued interest in graphene is the observation of a nonlinear response in it under the action of pulsed fields of the form (\ref{Etdz}) \cite{Bowlan:2014, Baudisch:2018}. Does the presented model based on the quantum kinetic equation reflect these properties of graphene? To answer this question, we study the spectral composition of internal currents shown in Figures \ref{fig:12} and \ref{fig:14}.

As a tool for studying time series, we use the discrete Fourier transform.
The figures (periodograms) below show the squares of the moduli of the coefficients of such a transformation
and reflect the contributions of various frequencies to the total field energy. 
The horizontal axis of the frequencies is calibrated in $THz$, the vertical axis has a logarithmic scale.
Since the absolute values of the periodogram depend, inter alia, on the length of the sample that we have, only relative values within one periodogram have a well-defined meaning.
The spectrum of the acting field plays the role of a reference point.

\begin{figure}[!h]
\begin{center}
	\includegraphics[width=0.48\textwidth]{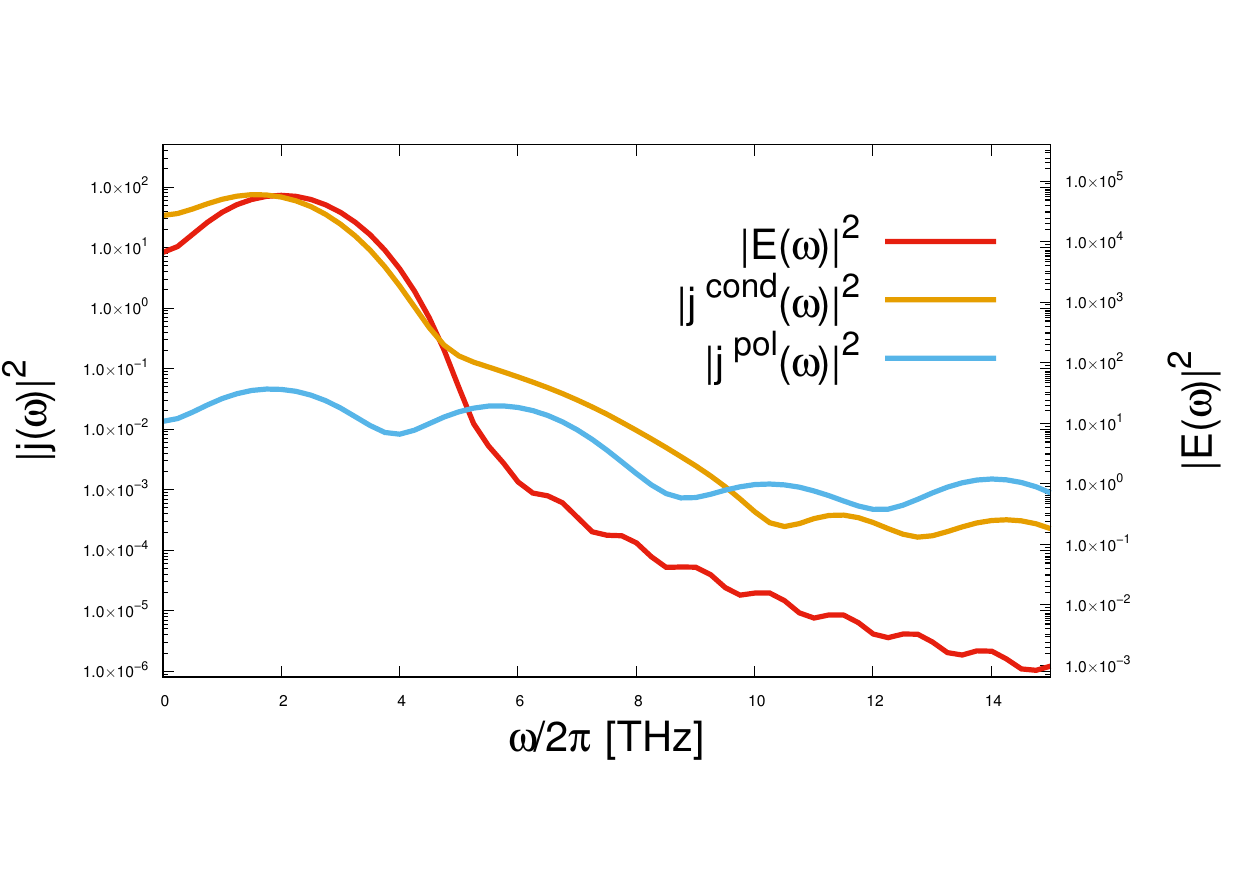}
	\end{center}
	\vspace{-6pt}
	\caption{Periodograms for the pulse of a periodic electric field (Figure \ref{fig:6}) and the currents generated by it (Figure \ref{fig:12}).		
	\label{fig:16}}
\end{figure}

However, in order to have an initial reference point, we first present in Figure \ref{fig:16} the result obtained from several discrete values of the external electric field shown in Figure \ref{fig:6}.
Since the pulse is very short, its spectrum is greatly broadened. 
Nevertheless, the maximum for the value of $2$~THz is quite visible, and outside it there is a smooth and fairly monotonic decrease in the intensity of the spectral components.
From this background we show the results of exactly the same processing of a sequence of conduction and polarization currents. The behavior of the spectrum of the conduction current is similar to the behavior of the spectrum of the acting field. However, the curve for the polarization current clearly contains local maxima for frequencies of $2$, $6$, $10$, and $14$~THz, which should be interpreted as the appearance of odd harmonics in the spectrum of the internal field with multiplicities of $3$, $5$, and $7$ with respect to the fundamental frequency.
If at the fundamental frequency the contribution to the field energy from the polarization current is about three orders of magnitude less then from the conduction current, at the third harmonic they are comparable, and at the fifth and seventh contributions of the polarization current are almost an order of magnitude larger.

\begin{figure}[!h]
\begin{center}
	\includegraphics[width=0.48\textwidth]{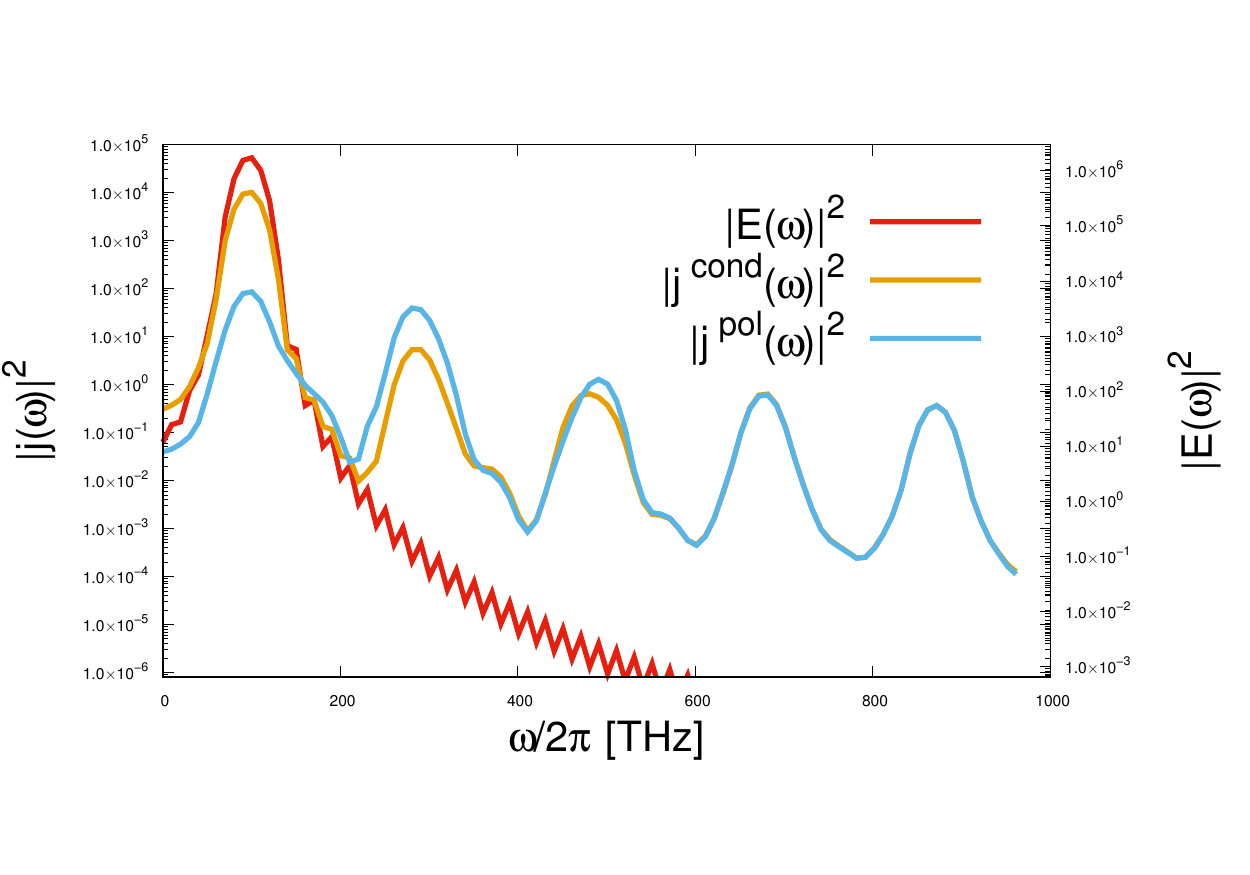}
\end{center}
\vspace{-2pt}
	\caption{Periodograms for the pulse of a periodic electric field (Figure \ref{fig:9}) and the currents generated by it (Figure \ref{fig:14}).
		\label{fig:18}}
\end{figure}

Let us now see how this looks in the case of a longer pulse, as depicted in Figure \ref{fig:9}. The spectrum of the latter is more localized. The spectrum of the pulse itself is shown in Figure \ref{fig:18}. The carrier frequency in this case is $96.7$~THz. Periodograms for currents are also shown in this figure.
In this case, the spectrum of both currents clearly shows the odd harmonics with numbers $ 3, 5, 7, 9, ... $ In the above figure, the upper frequency limit is $1000$ THz, but if we raise it, we will see harmonics of a higher order. At the carrier frequency, the contribution of the conduction current also dominates. However, already at the next, third, the contribution of the polarization current is about one and a half orders of magnitude greater. With increasing frequency, the contributions of the conduction and polarization currents become almost the same.
The appearance of odd harmonics was confirmed in experiments~\cite{Bowlan:2014, Baudisch:2018}.

The obtained results are in qualitative agreement with the existing experimental data, so that the suggested kinetic theory is surely verified.

\section{Conclusions \label{sect:c}}

In this work, we considered the application of the nonperturbative kinetic equation approach to describe the excitation of plasma oscillations in a graphene monolayer.
As examples the constant electric field as well as an electric field of short high-frequency pulses were considered. The~dependence of the induced conduction and polarization current characteristics on the pulse intensity, pulse duration, and polarization were investigated numerically for these examples.
The characteristics of secondary electromagnetic radiation resulting from the alternating currents was studied and a nonlinear response to the external electric field was found which characterizes graphene as an active medium.

A perspective direction of development of the kinetic theory of graphene is a generalization to the case of additionally accounting for the interaction with the quantized electromagnetic field.
One~can proceed from the analogy with the works \cite{Blaschke:2011, Smolyansky:2019f}, where such a generalization was performed in the standard QED on the basis of the Bogolyubov-Born-Green-Kirkwood-Yvon chain of equations in the single photon approximation.
Preliminary investigations of radiation on this basis met large difficulties~\cite{Smolyansky:2020}.
Analogous research in graphene would allow comparing the quasiclassical radiation (Section \ref{sect:pf}) with the quantum one and to understand deeper the situation in standard QED.
Considerable interest represents a study of cascade processes in graphene (e.g., \cite{Fedotov:2010, Nerush:2011} in standard QED) and also of spin~phenomena.

\section*{Acknowledgements}
We acknowledge B. Dora and R. Moessner for their stimulating interest 
in our work and for pointing out Ref. [6] to us. D.B. and N.G. are grateful to 
H. A. Sarkisyan for discussions about physics of low-dimensional systems. 
%\section*{Funding}
S.A.S. thanks for support Russian Science Foundation (Grant No. 19-12-00042).
%P.A.D. thanks for support by RFBR according to the research project 18-07-00778. 
S.A.S is grateful for support from the joint funding of Helmholtz Centres and JINR Dubna for his participation as a lecturer at the Helmholtz International Summer School in Dubna, July 2019, where the material of this article has been presented.

%%%%%%%%%%%%%%%%%%%%%%%%%%%%%%%%%%%%%%%%%%

%\reftitle{References}

%=====================================
% References, variant B: external bibliography
%=====================================
%\reftitle{References}
%\externalbibliography{yes}
%\bibliography{references}

%%%%%%%%%%%%%%%%%%%%%%%%%%%%%%%%%%%%%%%%%%
%% optional
%\sampleavailability{Samples of the compounds ...... are available from the authors.}

%% for journal Sci
%\reviewreports{\\
%Reviewer 1 comments and authors’ response\\
%Reviewer 2 comments and authors’ response\\
%Reviewer 3 comments and authors’ response
%}

%%%%%%%%%%%%%%%%%%%%%%%%%%%%%%%%%%%%%%%%%%
\end{document}